\documentclass[
reprint,
frontmatterverbose,preprintnumbers,nofootinbib,nobibnotes,amsmath,amssymb,aps,pra,rmp,prstab,rstper,floatfix,
]{revtex4-2}

\usepackage[colorlinks=true,
            linkcolor=red,      
            urlcolor=magenta,   
            citecolor=blue,     
            anchorcolor=red]{hyperref}
            
\usepackage{physics}
\usepackage{graphicx}
\usepackage{dcolumn}
\usepackage{bm}
\usepackage[mathlines]{lineno}
\usepackage{capt-of}
\AtBeginDocument{\addtolength{\textheight}{1cm}}

\makeatletter
\newcommand{\removewidetextbottomrule}{%
  \setbox\widetext@bot=\hbox{}%
}

\makeatother

\begin{document}
\title{Harnessing Post-selected von Neumann Measurements to Strengthen Entanglement
and Quantum Signatures in Coherent States}
\author{Janarbek Yuanbek}
\thanks{These authors contributed equally to this work.} \author{Zelalem Abebe Bekele}
\thanks{These authors contributed equally to this work.} 
\author{Zhi-qiang Liu}
\email{Corresponding author: lzq@semi.ac.cn}

\affiliation{Research and Development Center for Wide Bandgap Semiconductors, and}
\affiliation{State Key Laboratory of Semiconductor Physics and Chip Technologies,
Institute of Semiconductors, Chinese Academy of Sciences, Beijing
100083, China}
\affiliation{Center of Materials Science and Opto-Electronic Technology, University
of Chinese Academy of Sciences, Beijing 100049, China}
\date{\today}
\begin{abstract}
We investigate how post-selected von Neumann measurements modify the nonclassical properties of two-mode entangled coherent states (ECSs). From the conditional pointer state, we characterize squeezing, photon statistics, intermode correlations, entanglement, fidelity, and post-selection probability. Anomalous weak values with moderate coupling enhance squeezing, antibunching, sub-Poissonian statistics, intermode correlations, and bipartite entanglement, while fidelity captures the crossover from weak perturbation to strong distinguishability. These results demonstrate post-selection as a tunable approach to engineering ECSs.
\end{abstract}
\maketitle

\section{INTRODUCTION}

In the 1980s, Aharonov, Albert, and Vaidman introduced the concept of weak measurement (WM) \citep{PhysRevLett.60.1351}, marking a paradigm shift from the traditional projective (von Neumann) measurement framework that fundamentally reshaped the theoretical foundations of quantum measurement \citep{Hallaji2017,RevModPhys.86.307,4h4b-3xss}. WM explores the subtle interplay
between quantum systems and measuring apparatuses under minimal disturbance \citep{PhysRevResearch.7.013206}, enabling the extraction of weak values (WVs) that reveal otherwise inaccessible information about quantum states \citep{PhysRevD.40.2112,PhysRevA.41.11}. In contrast to strong measurements, which typically collapse the system into eigenstates \citep{Oriols2025}, WM preserves quantum coherence \citep{Bizzarri_2025} and, when combined with post-selection, facilitates anomalous weak value amplification (WVA) of specific observables \citep{394k-lj6s,PhysRevA.111.042425,vr7v-lwtb}, thereby providing a powerful tool for probing
non-classical phenomena \citep{g33l-rgdh,PhysRevLett.112.040406,PhysRevX.4.011031,10.1063/5.0342520}.

The applicability of this post-selection-based measurement framework for optimizing quantum states has been extensively investigated across diverse scenarios \citep{janarbek,DELIMABERNARDO2014194,janarbekPhysRevA.110.052611,Y-PhysRevA.105.022608,Y-PhysRevA.105.022210,Y-TUREK2023128663},
and many others see, e.g., Refs. \citep{70,RN2035,RN2066,2014Quantum,PhysRevA.85.012107}. However, a critical and unexplored challenge lies in understanding how post-selected von Neumann measurements (PVNMs) \citep{janarbek} influence the intrinsic properties of entangled quantum states,
particularly entangled coherent states (ECSs) \citep{PhysRevLett.107.083601,PhysRevA.45.6811,jeon2024experimental}. ECSs are non-Gaussian continuous-variable states \citep{Ahmad10062011,t5c6-byhh} 
formed from coherent superpositions distributed over distinct bosonic
modes
\citep{PhysRevResearch.7.023253,PhysRevApplied.20.044030,mgj7-t6d3}. 
Their simultaneous coherent-state and entangled structure gives rise to distinctive nonclassical properties \citep{PhysRevA.110.L010602,Israel19}, including quadrature squeezing, Einstein–Podolsky–Rosen (EPR) correlations \citep{Wu15}, and sub-Poissonian photon statistics \citep{ZENG2007162,BUKHARI20143788}, thereby providing valuable resources for quantum communication \citep{PhysRevA.65.042305,PhysRevA.68.042319,Sangouard:10,PhysRevLett.100.030503}, coherent-state quantum information processing \citep{doi:10.1126/science.1122858,PhysRevA.64.022302}, and quantum metrology \citep{PhysRevA.43.492,PhysRevLett.96.050503,PhysRevA.64.022313,PhysRevA.82.062325}.
Nevertheless, their sensitivity to measurement-induced disturbances remains an open question \citep{PhysRevA.70.020101}. Similarly, conventional strong measurements are known to degrade entanglement and non-classicality; the specific role of WMs, particularly those leveraging post-selection, has received comparatively limited exploration \citep{PhysRevX.4.011032,PhysRevA.77.062105}.

Previous studies have shown that post-selected WM can
amplify small optical signals and reshape nonclassical properties in
both single- and two-mode continuous-variable states
\citep{PhysRevA.81.033813,xldj-cz67,PhysRevA.108.L040202,PhysRevLett.134.080802}. Extending this paradigm to two-mode ECSs, however, introduces significant challenges \citep{PhysRevLett.113.030401}. The complex interplay between mode entanglement, quadrature squeezing, and post-selection effects necessitates a nuanced analysis \citep{Bernardo:14,Coto2017,Barral:16}. Although recent studies on continuous-variable systems have hinted at the potential for WM to reshape quantum statistics
and entanglement \citep{Yao2026,qmen-RevModPhys.81.865,qmen-Vedral2014QuantumE,qmen-Tavares2023QuantumE}, a comprehensive investigation into how PVNMs affect the core attributes of ECSs, including squeezing behavior, photon statistics and entanglement, remains conspicuously absent. To address this void, our study systematically builds upon the WM framework to clarify the revolutionary role of post-selection on ECSs, ultimately providing critical insights for quantum state engineering and the advancement of measurement-based quantum protocols.

The paper is structured as follows. Sec. \ref{sec:2} formulates the PVNMs framework for ECSs, deriving the final pointer state and its dependence on the measurement parameters. In Sec. \ref{sec:3}, examines quadrature and sum squeezing in the final pointer state to characterize the effect of PVNMs on ECSs. We then turn to Sec. \ref{sec:4} by an analysis of  correlations and non-classical photon statistics using second-order auto- and cross-correlation functions and the Mandel $Q$ parameter. Following this, Sec. \ref{sec:5} addresses entanglement through the Hillery-Zubairy and EPR correlations, evaluating the changes induced by the measurement process. In Sec. \ref{sec:6}, we quantify the fidelity between the initial and post-measurement states. Sec. \ref{sec:7} provides a discussion of this work. Sec. \ref{sec:8} summarizes the key findings and offers an outlook on promising directions for future research. Finally, Sec. \ref{sec:9}  provides details of the computational methodology. We adopt the natural unit system, in which $\hbar=1$.

\section{Fundamental principles \label{sec:2}}

In measurement theory, the total Hamiltonian is generally decomposed
into three fundamental components, each governing distinct physical
processes during the measurement interaction \citep{vonNeumann+2018,PhysRevA.76.044103,RevModPhys.86.307,PhysRevA.81.033819,Zhu2011}. These three components
are formally expressed as 
\begin{equation}
\hat{H}=\hat{H}_{\text{s}}+\hat{H}_{\text{p}}+\hat{H}_{\text{int}}.
\end{equation}

Here, $\hat{H}_{\text{s}}$, $\hat{H}_{\text{p}}$, and $\hat{H}_{\text{int}}$ denote the Hamiltonian of the measured system, the measuring apparatus (pointer), and their mutual interaction of the system and the apparatus, respectively. Within the framework of ideal (von Neumann) measurement
context \citep{vonNeumann+2018,70}, the specific forms of the Hamiltonians
described by $\hat{H}_{\text{s}}$ and $\hat{H}_{\text{p}}$ are assumed to be negligible over the measurement timescale; consequently, their explicit functional forms do not affect the readout statistics. Conversely, the interaction Hamiltonian $\hat{H}_{\text{int}}$ plays a pivotal role: it governs the measurement dynamics by defining the coupling strength and post-selection effects \citep{XU2024100518,10.1063/5.0230512}. In this study,
we adopt the following specific form of $\hat{H}_{\text{int}}$: 
\begin{equation}
\hat{H}_{\text{int}}=g(t)\hat{\sigma}_{x}\otimes\hat{P}_{x},\label{eq:introduction hamli}
\end{equation}
where $g(t)$ describes the time-dependent coupling strength between
the measured system and pointer. This $g(t)$ function is non-zero only over finite time intervals and satisfies the time-integrated condition of $\int^{t}_{0}g(\tau)d\tau=gt$, while $\hat{\sigma}_{x}$ is the measured observable, and $\hat{P}$ denotes the pointer is momentum operator (conjugate to position $\hat{X}$), which satisfies $[\hat{X},\hat{P}_{x}]=i$. Using the annihilation operator $\hat{a}$ and creation operator $\hat{a}^{\dagger}$, the momentum operator and position can be expressed as $\hat{P}_{x}=i(\hat{a}^{\dagger}-\hat{a})/2\sigma$, and $\hat{X}=\sigma(\hat{a}^{\dagger}+\hat{a})$ \citep{PhysRevA.41.1526}, here $\sigma=\sqrt{1/2m\omega}$ represents the Gaussian ground-state width of the pointer, which is determined by its mass $m$ and oscillation frequency $\mathrm{\omega}$. The corresponding experimental configuration is illustrated in Fig. \ref{fig:weak-measurement}. Thus, the initial composite state of the system is given by:

\begin{equation}
\vert\Psi_{\text{in}}\rangle=\vert\psi_{\text{i}}\rangle\otimes\vert\phi\rangle,\label{int}
\end{equation}
where $\vert\psi_{\text{i}}\rangle=\cos(\theta/2)\vert H\rangle+e^{i\delta}\sin(\theta/2)\vert V\rangle$ is the initial state of the measured system and $\vert\phi\rangle$ denotes the pointer.
Here, $\delta\in[0,2\pi]$ and $\theta\in[0,\pi)$ are dimensionless azimuthal and polar angles \citep{Ryu2019}, respectively, where $\delta$ specifies the relative phase and $\theta$ determines the relative populations of the $\vert H\rangle$ and $\vert V\rangle$ components of the measured system \citep{PhysRevLett.94.220405}. Assuming that the ECSs, represented as: 
\begin{align}
\vert\phi\rangle & =\mathcal{N}(|\alpha\rangle_{\text{a}}\vert0\rangle_{\text{b}}+|0\rangle_{\text{a}}\vert\alpha\rangle_{\text{b}}),\label{pointer-state}
\end{align}
where $\vert0\rangle$ and $\vert\alpha\rangle$ denote the Fock vacuum and coherent states in spatial modes $\text{a}$ and $\text{b}$ \citep{PhysRevA.81.033819,PhysRevLett.107.083601,PhysRevA.64.054102,PhysRevA.76.031806,PhysRevA.79.033822},
respectively, and $\mathcal{N}=[2(1+e^{-|\alpha|^{2}})]^{-1/2}$ is the normalization factor,
here $\alpha=re^{i\mu}$ while $r$ is coherent state amplitude and $\mu$ is its phase, with $ r \in[0,\infty) $ and $ \mu \in[0,2\pi] $. 
The polarization pre-selection may be implemented using suitable wave plates, whereas the two-mode ECSs in Eq. (\ref{pointer-state}) are assumed to be prepared independently by an appropriate non-Gaussian optical-state-generation protocol.
The joint state expressed in Eq. (\ref{int}) evolves by means of the time evolution operator, $U(t)=\exp[-i\int^{t}_{0}H_{\text{int}}d\tau]$. Thus, the  $\vert\Psi\rangle=U(t)\vert\psi_{\text{i}}\rangle\otimes\vert\phi\rangle$ becomes: 
\begin{align}
\vert\Psi\rangle & =\frac{1}{2}\left[J_{+}\hat{D}_{\text{a}}\left(\frac{s}{2}\right)+J_{-}\hat{D}^{\dagger}_{\text{a}}\left(\frac{s}{2}\right)\right]\vert\psi_{i}\rangle\otimes\vert\phi\rangle,
\end{align}
here $J_{\pm}=\mathbb{I}\pm\hat{\sigma}_{x}$, with $\mathbb{I}$ is the $2\times2$ unit matrix operator, $\hat{D}_{\text{a}}(\pm s/2)=\exp[s(\pm \hat{a}^{\dagger}\mp \hat{a})/2]$
the \emph{displacement} operator, and $s=gt/\sigma$ parametrizes the
measurement strength (for WM $s$ is small, i.e., $s\ll1$) corresponds to weak coupling,
while strong measurements (large $s\gg1$) indicate strong coupling.
We thus term $s$ the transition parameter of the measurement model,
then the pointer state reads $\vert\tilde{\Phi}\rangle=\langle\psi_{\text{f}}\vert\Psi\rangle$, where $\vert\psi_{\text{f}}\rangle=\vert H\rangle$ denotes the post-selected state of the measured system.
The state $\vert\tilde{\Phi}\rangle$ is not normalized, the above
state $\vert\tilde{\Phi}\rangle$ expressed as: 
\begin{align}
\vert\tilde{\Phi}\rangle & =\frac{\langle\psi_{\text{f}}\vert\psi_{i}\rangle}{2}\bigg[t_{+}\hat{D}_{\text{a}}\left(\frac{s}{2}\right)+t_{-}\hat{D}^{\dagger}_{\text{a}}\left(\frac{s}{2}\right)\bigg] \vert\phi\rangle,\label{eq:not-normalization}
\end{align}
here $t_{\pm}=1\pm\langle\hat{\sigma}_{x}\rangle_{w}$, by imposing
the normalization condition $\langle\psi_{\text{i}}\vert\psi_{\text{i}}\rangle=\langle\psi_{\text{f}}\vert\psi_{\text{f}}\rangle=1$,
the overlap $\langle\psi_{\text{f}}\vert\psi_{\text{i}}\rangle=\cos(\theta/2)$
naturally satisfies the constraints on the WV parameter $\theta$.
We proceed to impose normalization on the derived equation $\vert\Phi\rangle=\vert\tilde{\Phi}\rangle/\sqrt{P_{\rm suc}}=\mathcal K \vert\tilde{\Phi}\rangle$. Here, $\mathcal K=1/\sqrt{P_{\text{suc}}}$ and $P_{\rm suc}=\langle\tilde{\Phi}|\tilde{\Phi}\rangle$
can also characterize the probability of successful post-selection
of the final pointer state $\vert\Phi\rangle$ and we can define as:

\begin{figure}[t]
\begin{centering}
\includegraphics[width=\linewidth]{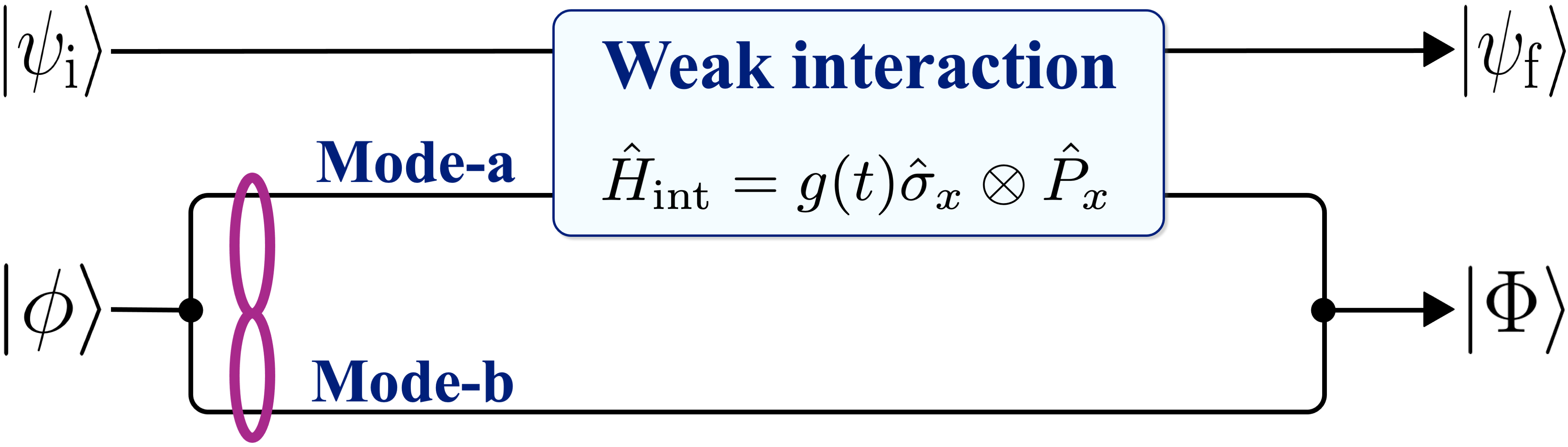}
\par\end{centering}
\caption{Schematic of the PVNMs protocol for preparing the conditional pointer state $|\Phi\rangle$ in mode-$a$. The system and pointer are initially prepared in the states $|\psi_{\text{i}}\rangle$ and $|\phi\rangle$, respectively, coupled through a weak interaction, and subsequently post-selected by projecting the system onto the final state $|\psi_{\text{f}}\rangle$.
\label{fig:weak-measurement}}
\end{figure}

\begin{align}
P_{\rm suc} =\frac{\cos^{2}(\theta/2)}{2} \{1+\vert\langle\hat{\sigma}_{x}\rangle_{w}\vert^{2}+\mathrm{Re}[t_{-}t^{*}_{+} \beta ]\},\label{eq:successful post-selection}
\end{align}
where
$
\beta = \langle\phi\vert\hat{D}_{\text{a}}(s)\vert\phi\rangle =\mathcal{N}^2\{
(1+e^{-2is\,\operatorname{Im}\alpha})e^{-s^2/2}
+e^{-|\alpha|^2/2}e^{-is\,\operatorname{Im}\alpha}
[
e^{-|\alpha-s|^2/2}
+e^{-|\alpha+s|^2/2}
]\}.
$
Building directly upon the expression above, the main objective of this study is to examine the influence of post-selected measurement on the intrinsic properties of the ECSs, thereby yielding the final pointer state given by: 

\begin{equation}
\vert\Phi\rangle=\frac{\kappa}{2}\left[t_{+}\hat{D}_{\text{a}}\left(\frac{s}{2}\right)+t_{-}\hat{D}^{\dagger}_{\text{a}}\left(\frac{s}{2}\right)\right] \vert\phi\rangle,\label{eq:Psi}
\end{equation}
where 
$\kappa  =\sqrt{2}/\left[1+\vert\langle\hat{\sigma_{x}}\rangle_{w}\vert^{2}+(1-\vert\langle\hat{\sigma}_{x}\rangle_{w}\vert^{2})\mathrm{Re}[\beta]\right]^{1/2}
$ is the normalization coefficient
and $\langle\hat{\sigma}_{x}\rangle_{w}$ is the WV of the system
observable $\hat{\sigma}_{x}$ defined by 
\begin{equation}
\langle\hat{\sigma}_{x}\rangle_{w}=\frac{\langle\psi_{f}\vert\hat{\sigma}_{x}\vert\psi_{i}\rangle}{\langle\psi_{f}\vert\psi_{i}\rangle}=e^{i\delta}\tan\frac{\theta}{2}.\label{eq:8}
\end{equation}

The WV is obtained via a measurement with a post-selection probability
of $P_{0}=\vert\langle\psi_{f}\vert\psi_{i}\rangle\vert^{2}=\cos^{2}(\theta/2)$ \citep{PhysRevLett.66.1107,science.1152697}.
In post-selected WM, the value of the measured observable is given by WV. Its real ($\mathrm{Re}$) and imaginary ($\mathrm{Im}$)
components are extracted from the position and momentum shifts of the measuring device's pointer, respectively. The Fourier transform provides the conversion between position
and momentum spaces \citep{Gao:25,Maier2026}. Next, we investigate the effects of a PVNM scheme and the WV of the measured observable on the intrinsic properties of the ECSs.

\section{EFFECTS ON SQUEEZING\label{sec:3}}

This section examines how PVNMs influence the squeezing properties of ECSs, considering both quadrature and sum squeezing.

\subsection{Quadrature squeezing\label{sec:3-1}}

In continuous-variable quantum optics, quadrature squeezing and non-Gaussian
entanglement is a fundamentally non-classical property of light fields \citep{WANG2015108,Turek2021}, essential for quantum computation and communication protocols \citep{PhysRevA.28.1560,PhysRevLett.130.090602,mgj7-t6d3,PhysRevD.23.1693}. Here, we investigate how PVNMs affect the quadrature squeezing
of the state $|\Phi\rangle$. We define the two-mode quadrature operators
as $\hat{F}_{1}=(\hat{a}+\hat{b}+\hat{a}^{\dagger}+\hat{b}^{\dagger})/\sqrt{8}$
and $\hat{F}_{2}=(\hat{a}+\hat{b}-\hat{a}^{\dagger}-\hat{b}^{\dagger})/\sqrt{8}i$
\citep{SCHNABEL20171}. These operators follow the commutation relation
$[\hat{F}_{1},\hat{F}_{2}]=i/2$. The uncertainty relation governing
their fluctuations is $\Delta^{2}F_{1}\Delta^{2}F_{2}\geq1/16,$ where
$\Delta^{2}F_{i}=\langle\text{\ensuremath{\hat{F}^{2}_{i}}}\rangle-\langle\hat{F}_{i}\rangle^{2}$,
for $i=1,2$. Like in single-mode systems, two-mode squeezing happens
when any variance $\triangle^{2}F_{i}$ drops below the shot noise
level of 0.25. This occurs either when uncorrelated modes are $i=1$ and $i=2$ of $F_{i}$ individually squeezed, or when non-classical correlations (such as entanglement) exist between modes. We define the squeezing parameter for the quadrature squeezing of the i-th component in $\vert\Phi\rangle$ as $Q_{i}=\Delta^{2}F_{i}-0.25.$ The squeezing parameter $Q_{i}$
satisfies $Q_{i}\geq-0.25$. Quadrature squeezing occurs for the $i$-th component when $-0.25\leq Q_{i}<0$.

For the final state $|\Phi\rangle$, we obtain the mode-specific quadrature squeezing parameters $Q_{1}$ and $Q_{2}$ as:

\begin{align}
Q_{1}= & \frac{1}{4}\left[\langle\hat{a}^{\dagger}\hat{a}\rangle+\langle\hat{b}^{\dagger}\hat{b}\rangle+\operatorname{Re}[\langle\hat{a}^{2}\rangle+\langle\hat{b}^{2}\rangle]\right]+\frac{1}{2}\operatorname{Re}[\langle\hat{a}\hat{b}\rangle]\nonumber \\
 & +\frac{1}{2}\operatorname{Re}[\langle\hat{a}^{\dagger}\hat{b}\rangle]-\frac{1}{2}\left[\operatorname{Re}[\langle\hat{a}\rangle]+\operatorname{Re}[\langle\hat{b}\rangle]\right]^{2},
\\
Q_{2}= & \frac{1}{4}\left[\langle\hat{a}^{\dagger}\hat{a}\rangle+\langle\hat{b}^{\dagger}\hat{b}\rangle-\operatorname{Re}[\langle\hat{a}^{2}\rangle+\langle\hat{b}^{2}\rangle]\right]-\frac{1}{2}\operatorname{Re}[\langle\hat{a}\hat{b}\rangle]\nonumber \\
 & +\frac{1}{2}\operatorname{Re}[\langle\hat{a}^{\dagger}\hat{b}\rangle]-\frac{1}{2}\left[\operatorname{Im}[\langle\hat{a}\rangle]+\operatorname{Im}[\langle\hat{b}\rangle]\right]^{2}.
\end{align}

\onecolumngrid
\vspace*{\fill}
\begin{center}
\setlength{\tabcolsep}{0.5pt}
\begin{tabular}{@{}cc@{}}
\includegraphics[width=0.45\textwidth]{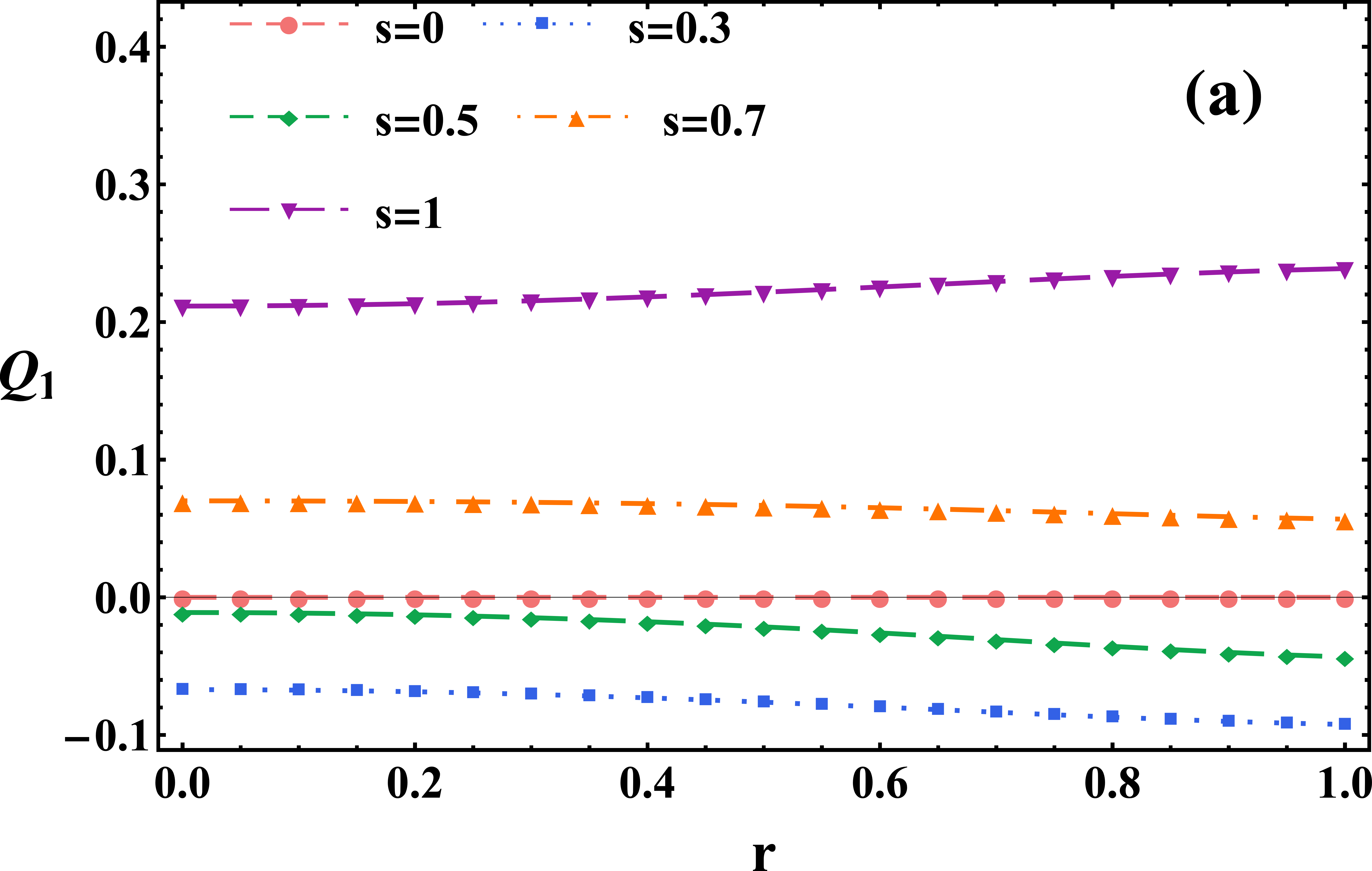} &
\includegraphics[width=0.466\textwidth]{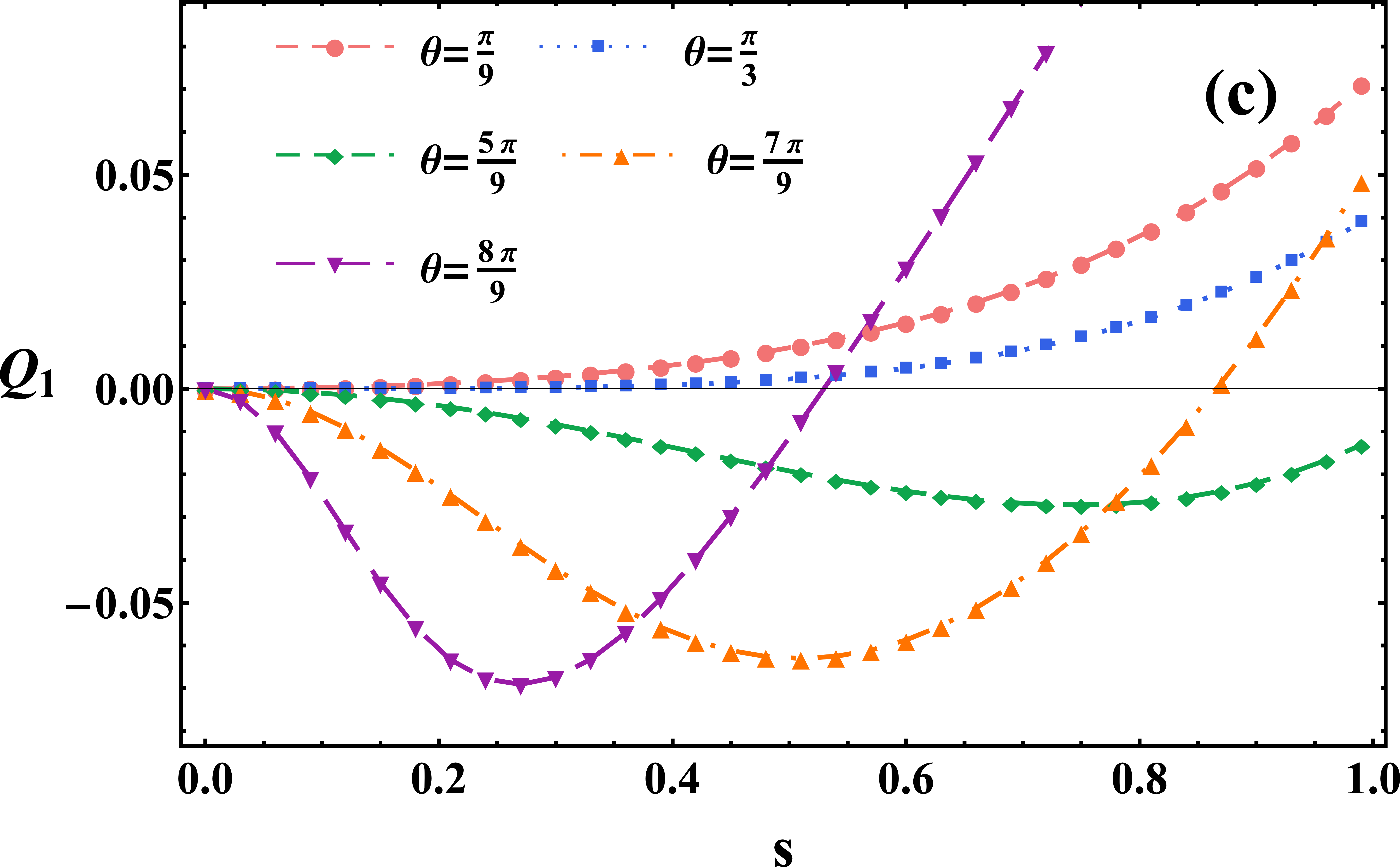} \\[-2.5pt]
\includegraphics[width=0.45\textwidth]{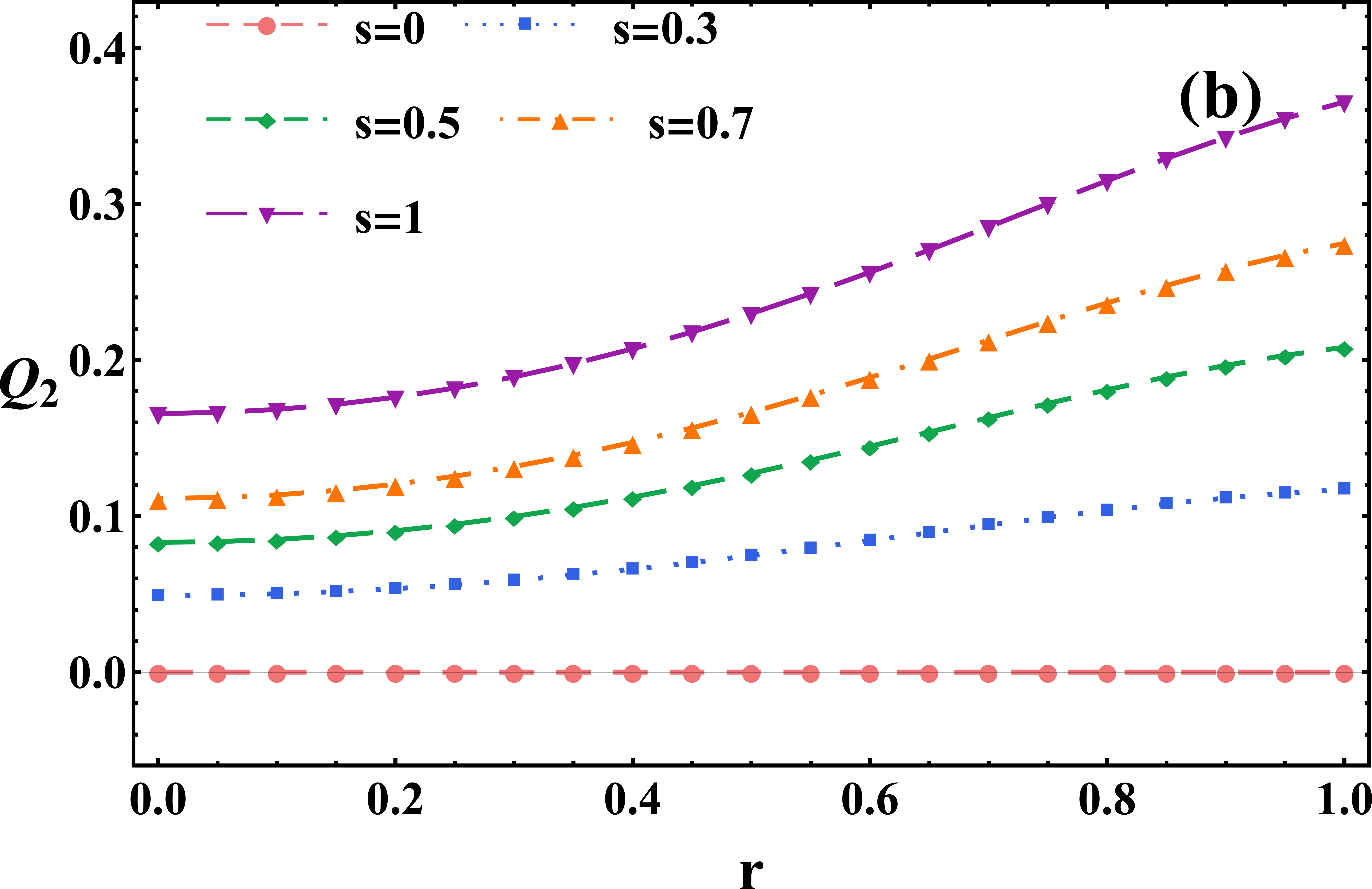} &
\includegraphics[width=0.467\textwidth]{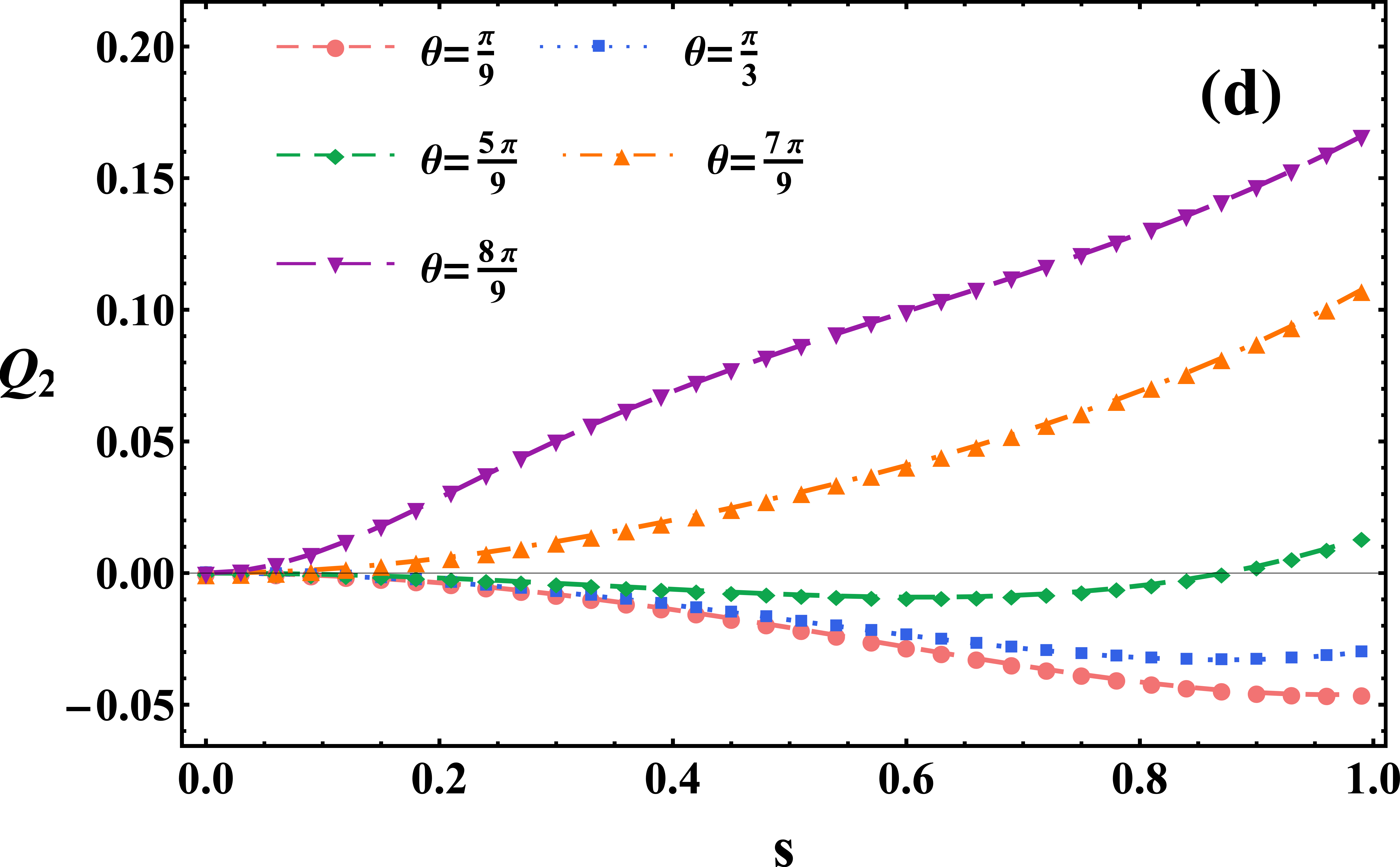}
\end{tabular}
\end{center}
\vspace{-12pt}
\captionof{figure}{Quadrature squeezing parameters $Q_{1}$ and $Q_{2}$. (a) and (b) $Q_{1}$ and $Q_{2}$ as functions of $r$ for different coupling strengths $s$, with the WV fixed at $\theta=8\pi/9$, (c) and (d) $Q_{1}$ and $Q_{2}$ as functions of the coupling strength $s$ for different WVs, with $r=0.1$ fixed. Here, $\delta=\mu=0$.\label{fig:Quadratic-squeezing}}
\vspace{-4pt}
\newpage
\twocolumngrid

Here, $\langle\cdots\rangle$ denotes quantum expectation values for
the state $\vert\Phi\rangle$, with analytic expressions provided
in Appendix \ref{sec:A1}. Setting $s=0$ recovers the inherent
squeezing of the ECSs ($\vert\phi\rangle$), characterized by $Q_{1,\phi}=Q_{2,\phi}=0$. To elucidate the impact of PVNMs on quadrature squeezing in $\vert\Phi\rangle$, we analyze numerical results as shown in Fig. \ref{fig:Quadratic-squeezing}. Figures \ref{fig:Quadratic-squeezing}(a) and (b) display $Q_{1}$ and $Q_{2}$ as functions of $r$ for distinct coupling strengths $s$, with the WV fixed at $\langle\hat{\sigma}_{x}\rangle_{w}=5.671$. Within the WM regime, the ECSs exhibit no orthogonal quadrature squeezing across any parameter range (red curves, $s=0$). For $s=0.3$ and $s=0.5$, $Q_{1}$ squeezing exceeds that of the initial state, demonstrating weak-value amplification [Fig. \ref{fig:Quadratic-squeezing}(a)], Conversely, $Q_{2}$ shows no enhancement [Fig. \ref{fig:Quadratic-squeezing}(b)], consistent with the principle that enhanced squeezing along one axis leads to reduced squeezing along the orthogonal axis. Figs. \ref{fig:Quadratic-squeezing}(c) and (d) plot $Q_{1}$ and $Q_{2}$ against $s$ for varying WVs at $r=0.1$. Large WVs induce $Q_{1}$ squeezing within a narrow coupling range ( $0<s<0.36$)
[Fig. \ref{fig:Quadratic-squeezing}(c)], while attenuating
$Q_{2}$ squeezing [Fig. \ref{fig:Quadratic-squeezing}(d)].
When $\delta=\mu=0$, increased $s$ in the strong measurement regime
suppresses the squeezing enhancement for both $Q_{1}$ and $Q_{2}$. Nevertheless, these results confirm that post-selected WVs effectively enhance ECSs squeezing. To further verify this conclusion, we next investigate the sum squeezing properties.

\subsection{Sum squeezing \label{sec:3-2}}

Sum squeezing characterizes two mode non-classicality by reduced fluctuations
in the quadrature $V_{\Theta}$ for modes $a$ and $b$ \citep{PhysRevA.40.3147},
defined as $V_{\Theta}=(e^{i\Theta}\hat{a}^{\dagger}\hat{b}^{\dagger}+e^{-i\Theta}\hat{a}\hat{b})/2$,
where $\Theta$ is the phase angle of $V_{\Theta}$ in the complex
plane \citep{NGUYENBAAN199934,PhysRevLett.112.070402,REN2019106,PhysRevA.40.3147}.
Sum squeezing occurs when $\langle(\Delta V_{\Theta})^{2}\rangle<\langle\hat{a}^{\dagger}\hat{a}+\hat{b}^{\dagger}\hat{b}+1\rangle/4,$ with
the degree of sum squeezing is defined as:

\begin{equation}
S=\frac{4\left\langle (\Delta V_{\Theta})^{2}\right\rangle }{\langle\hat{a}^{\dagger}\hat{a}+\hat{b}^{\dagger}\hat{b}+1\rangle}-1.\label{eq:sum squeezing}
\end{equation}

Sum squeezing manifests when $S<0$, with its theoretical minimum at $S=-1$. The degree of sum squeezing increases monotonically as $S$ approaches this lower bound. Substituting the two-mode quadrature observable $V_{\Theta}$ into Eq. (\ref{eq:sum squeezing}) yields an expression in terms of normally ordered operators

\begin{align*}
S  =\frac{2\big[\operatorname{Re}[e^{-2i\Theta}\langle\hat{a}^{2}\hat{b}^{2}\rangle]-2(\operatorname{Re}[e^{-i\Theta}\langle\hat{a}\hat{b}\rangle])^{2}+\langle\hat{a}^{\dagger}\hat{a}\hat{b}^{\dagger}\hat{b}\rangle\big]}{\langle\hat{a}^{\dagger}\hat{a}\rangle+\langle\hat{b}^{\dagger}\hat{b}\rangle+1}.
\end{align*}

\begin{figure}[t]
\begin{centering}
\includegraphics[width=0.47\textwidth]{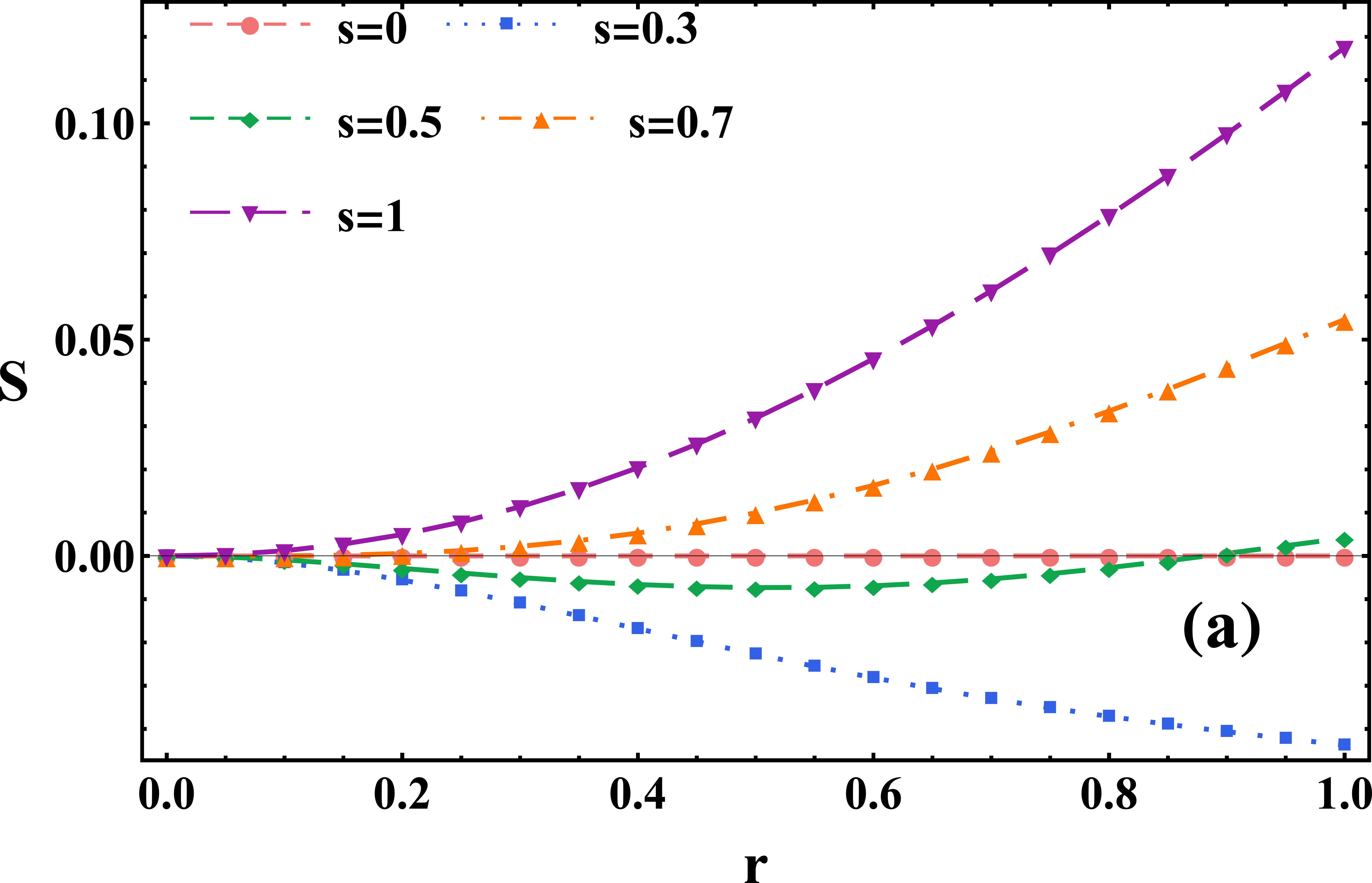}
\includegraphics[width=0.485\textwidth]{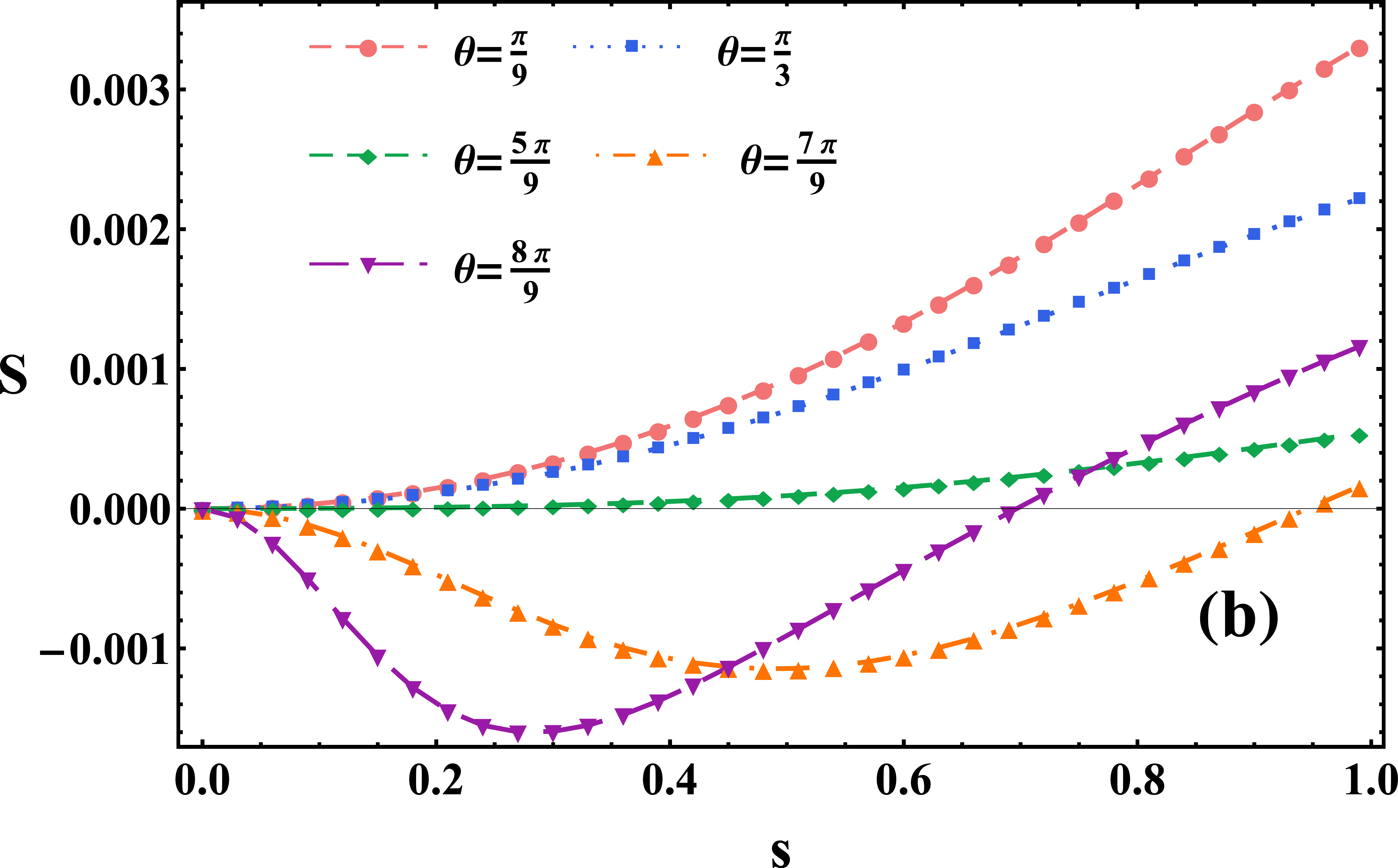}
\par\end{centering}
\caption{(a) $S$ as functions of $r$ for different coupling strengths $s$,
with the WV fixed at $\theta=8\pi/9$, (b) $S$ as functions
of the coupling strength $s$ for different WVs, with fixed
$r=0.1$ . Here, $\delta=\Theta=\mu=0$.\label{fig:Sum-sq}}
\end{figure}

Here, $\langle\cdots\rangle$ denotes quantum expectation values (Appendix
\ref{sec:A1}). Setting $\mathrm{s=0}$ recovers the inherent sum
squeezing ($S=0$) of the ECSs. The
behavior of the squeezing parameter $S$ under controlled conditions is
presented in Fig. \ref{fig:Sum-sq}. In Fig. \ref{fig:Sum-sq} (a)
$S$ is plotted as a function of rfor distinct coupling strengths
$s$, with the WVs fixed at $\langle\hat{\sigma}_{x}\rangle_{w}=5.671$.
Adopting the same analytical methodology applied to Fig. \ref{fig:Quadratic-squeezing}, it is found that the squeezing effect induced by $s=0.3$ exceeds that of the initial state ($s=0$). Figure \ref{fig:Sum-sq} (b) illustrates the relationship between $S$ and $s$ for varying WVs at $r=0.1$, emphasizing that substantial squeezing emerges within the coupling parameter regime and corresponds to larger WVs. The numerical results from Figs. \ref{fig:Quadratic-squeezing} and \ref{fig:Sum-sq} indicate that PVNMs enhance the squeezing of ECSs. However, this effect may not inherently express a quantum property of the field. Therefore, additional criteria are necessary to fully validate the quantum enhancement. 

\section{The effects on quantum statistics\label{sec:4}}

\subsection{Second-order correlation function\label{subsec:4-1}}

\begin{figure}
\begin{centering}
\includegraphics[width=0.48\textwidth]{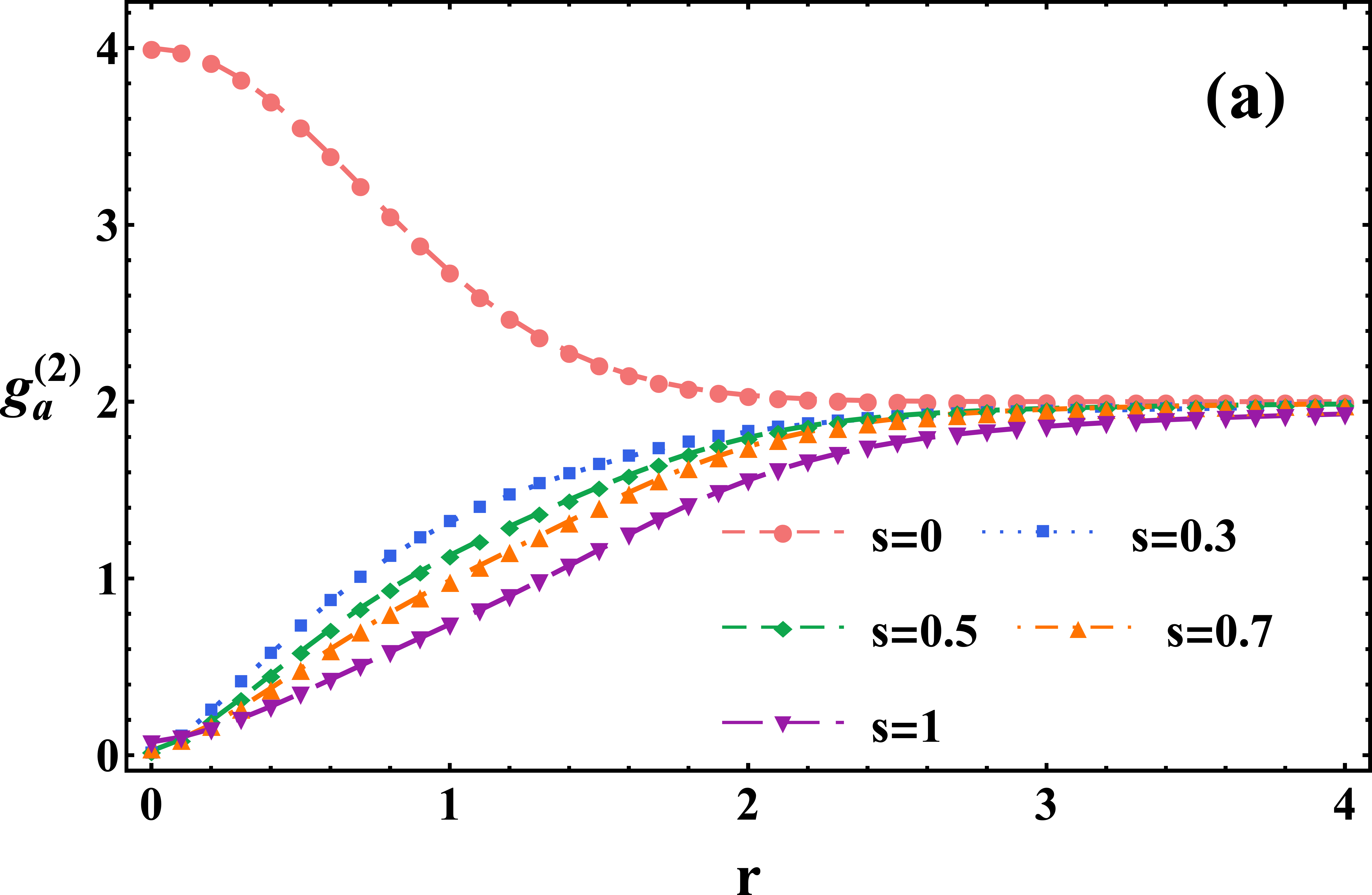}
\\
\includegraphics[width=0.48\textwidth]{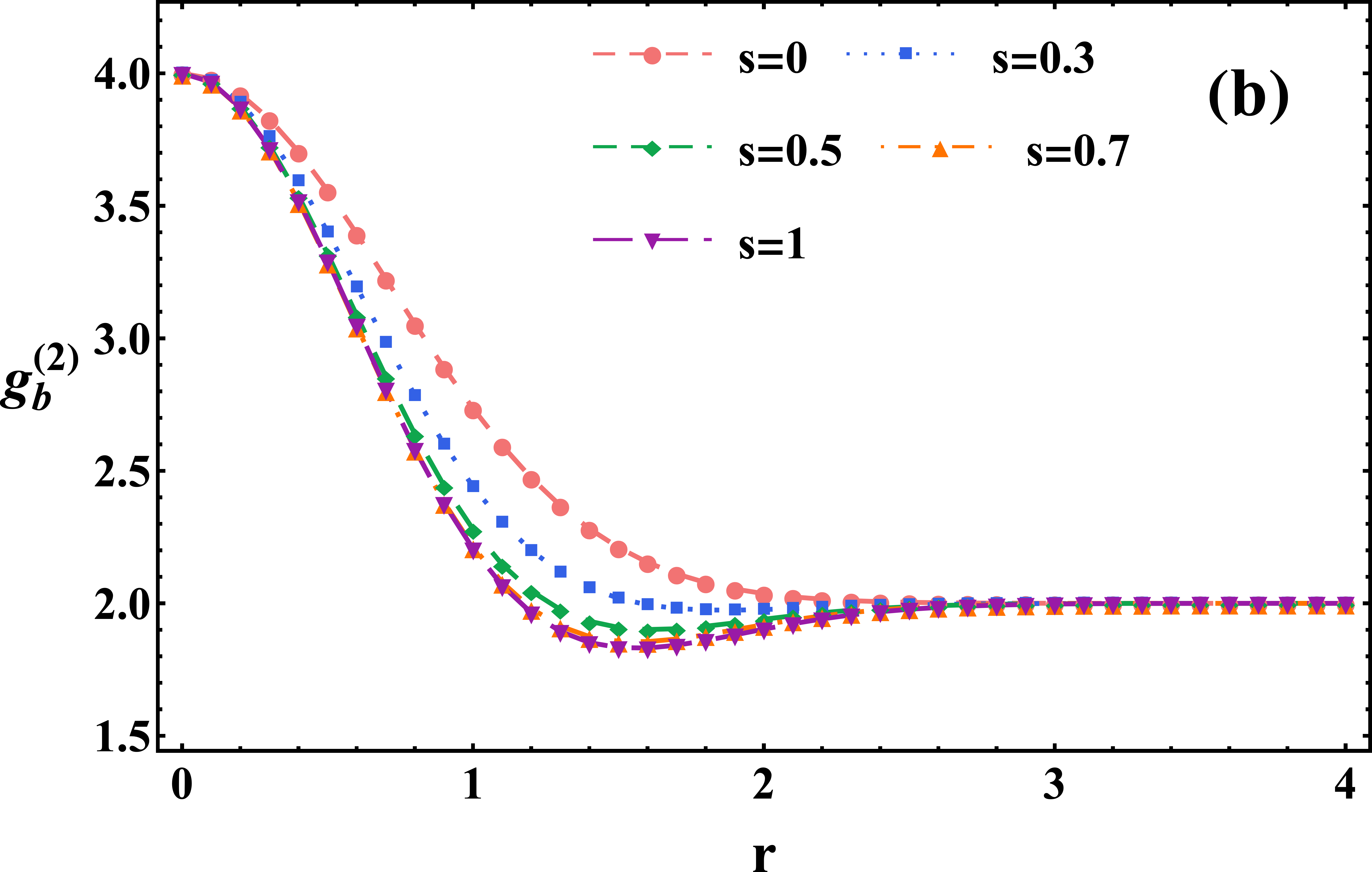}
\par\end{centering}
\caption{(a) and (b) as $g^{(2)}_{a}(0)$ and $g^{(2)}_{b}(0)$ functions of
$r$ for different coupling strengths $s$, with the WV fixed
at $\theta=8\pi/9$. Here, $\delta=\mu=0$.\label{fig:g2-ab}}
\end{figure}

The zero-delay second-order correlation (SOC) function can conveniently characterize the statistical properties of the radiation field \citep{lykp-833h}. A value in the range $0\le g^{(2)}_i\le 1$ ($i=a,b$) indicates sub-Poissonian photon statistics, which is widely recognized as a marker of the field's non-classical nature \citep{Agarwal2013}. For the quantum states investigated in this work, the zero-delay SOC functions for each of the two modes are respectively given as:  
\begin{align}
g^{(2)}_{a,\Phi}(0)  =\frac{\langle\hat{a}^{\dagger2}\hat{a}^{2}\rangle}{\langle\hat{a}^{\dagger}\hat{a}\rangle^{2}}, 
\quad 
g^{(2)}_{b,\Phi}(0)  =\frac{\langle\hat{b}^{\dagger2}\hat{b}^{2}\rangle}{\langle\hat{b}^{\dagger}\hat{b}\rangle^{2}}.
\end{align}

Here, $\langle\cdots\rangle$ denotes quantum expectation values for the state $\vert\Phi\rangle$, with analytic expressions detailed in Appendix \ref{sec:A1}. 
For the ECSs, this yields $g^{(2)}_{a,\phi}(0)$=$g^{(2)}_{b,\phi}(0)=\mathcal{N}^{-2}>1$. The $r$ dependence of $g^{(2)}_{a,\Phi}(0)$ and $g^{(2)}_{b,\Phi}(0)$ is examined in Fig. \ref{fig:g2-ab}. Figs. \ref{fig:g2-ab}(a) and (b) plot these functions against $r$ for distinct coupling strengths $s$, with the WV fixed at $\langle\hat{\sigma}_{x}\rangle_{w}=5.671$. When the absence of interaction ($s=0$), the two modes exhibit mode-dependent photon statistics at small $r$, mode a shows anti-bunching ($g^{(2)}_{a,\Phi}(0)<1$), indicating sub-Poissonian statistics, while mode b displays superbunching
($g^{(2)}_{b,\Phi}(0)>1$). The observed features are primarily attributable to the PVNMs, which significantly modify the statistical properties of photons, particularly in the regime where the WV is large. As the coupling coefficient $s$ increases, mode a exhibits clear anti-bunching characteristics, a typical feature of non-classical light sources (e.g., single-photon emitters) that cannot be explained by classical light wave theory. To rigorously corroborate the conclusions drawn from the second-order correlation function analysis, we further employ the Mandel $Q$ parameter as a complementary verification tool. 

\subsection{Mandel $Q$ parameter\label{subsec:4-2}}

\begin{figure}[t]
\begin{centering}
\includegraphics[width=1\columnwidth]{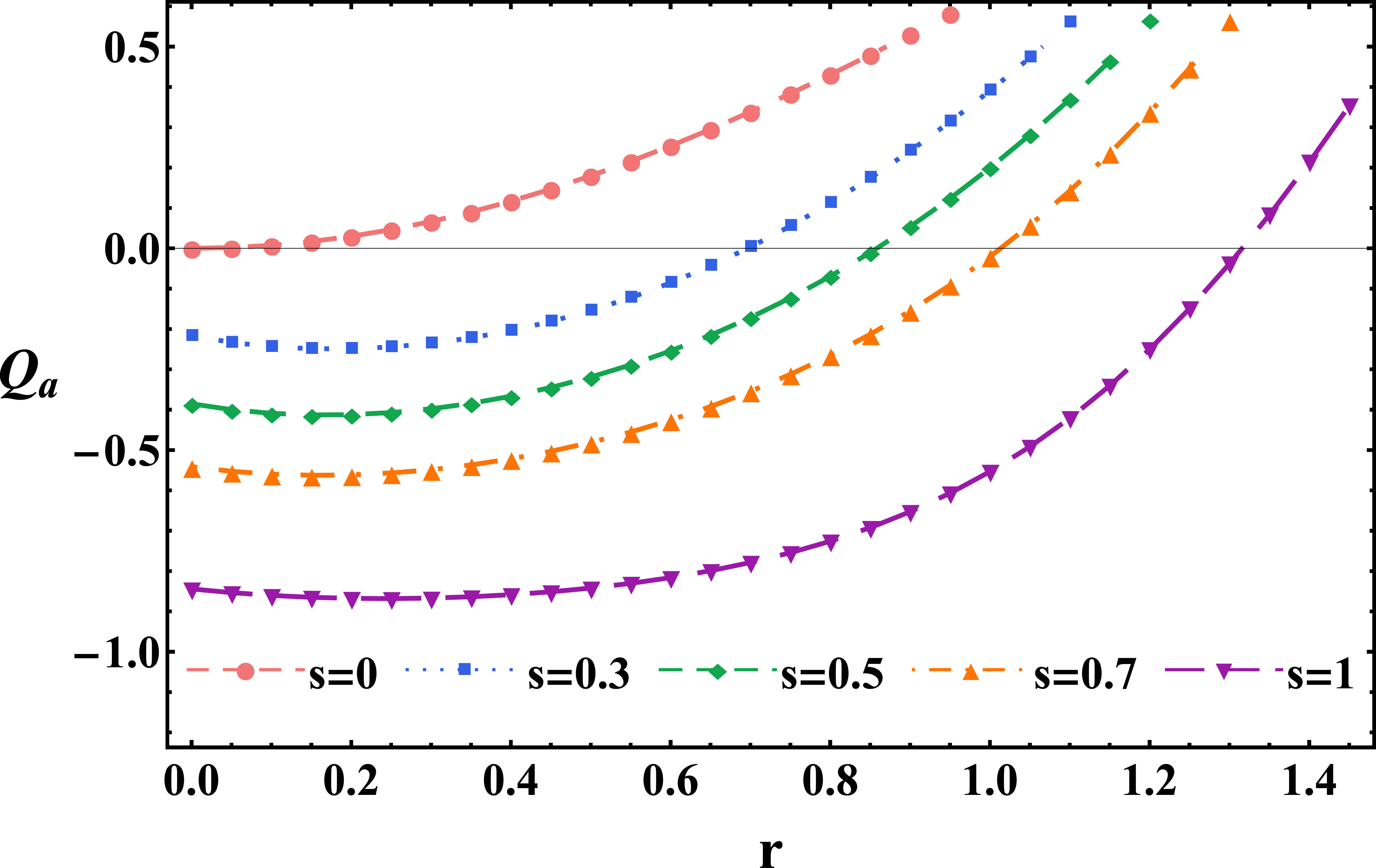}
\par\end{centering}
\caption{Mandel $Q$ factor as functions of $r$ for different coupling strengths
$s$, with the WV fixed at $\theta=8\pi/9$. Here, $\delta=\mu=0$.\label{fig:Mandel--factor}}
\end{figure}

Photon-number statistics, quantified by the Mandel $Q$ (MQ) parameter, classify optical fields as sub-Poissonian, Poissonian, or super-Poissonian, with ($Q<0$) signaling nonclassicality. In this subsection, we employ the MQ parameter to investigate the effects of PVNMs on the photon-number statistics of ECSs,
which is defined as follows \citep{Agarwal2013}:

\begin{equation}
Q_{a,\Phi}=\frac{\langle\hat{a}^{\dagger2}\hat{a}^{2}\rangle-\langle\hat{a}^{\dagger}\hat{a}\rangle^{2}}{\langle\hat{a}^{\dagger}\hat{a}\rangle}.\label{eq:20-1}
\end{equation}

Below, we focus on the condition $Q_{a,\Phi}<0$ for the generated state. We obtain the explicit expression of $Q_{a,\Phi}$ for the state $\vert\Phi\rangle$ by substituting the relevant expectation values listed in Appendix \ref{sec:A1}. 
When no measurement is applied $s=0$, we find that the MQ parameter for the ECSs is $Q_{a,\phi}=r^{2}(1-\mathcal{N}^{2})>0$. Through PVNMs under weak coupling ($s>0$), in Fig. \ref{fig:Mandel--factor} shows the MQ parameter versus $r$ for distinct coupling strengths $s$, with the larger WVs $\langle\hat{\sigma}_{x}\rangle_{w}=5.671$, $Q_{a}$ becomes negative within the range $0<r\leqslant1.35$. This indicates that post-selected WM is an anomalous WVs transform that shifts the quantum statistics of ECSs from super-Poissonian to sub-Poissonian, thereby enhancing their non-classicality. This result can also be observed through the zero-time-delay second-order cross-correlation function, which will be discussed in detail in the next subsection.

\subsection{Second-order cross-correlation function\label{subsec:4-3}}

Here we calculate the second-order cross-correlation (SOCC) function for ECSs prepared via PVNMs using the state $\vert\Phi\rangle$. The SOCC function for a two-mode field
is defined as\citep{REN2019106,GIRI2017140} 

\begin{equation}
g^{(2)}_{a,b}=\frac{\langle\hat{a}^{\dagger}\hat{a}\hat{b}^{\dagger}\hat{b}\rangle}{\langle\hat{a}^{\dagger}\hat{a}\rangle\langle\hat{b}^{\dagger}\hat{b}\rangle}.\label{eq:20}
\end{equation}

\begin{figure}
\begin{centering}
\includegraphics[width=1\columnwidth]{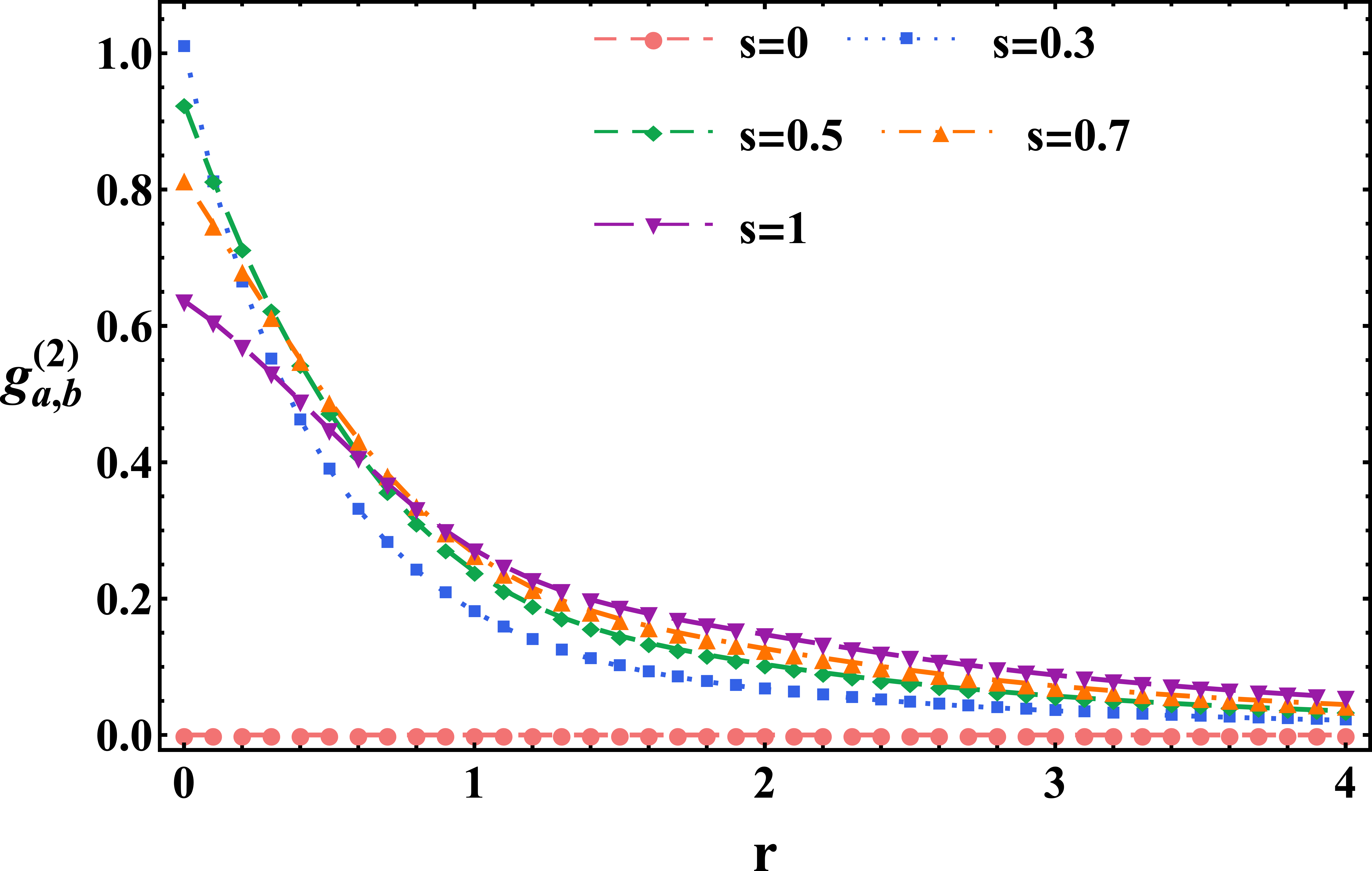}
\par\end{centering}
\caption{SOCC function $g^{(2)}_{a,b}$ as functions of $r$ for different
coupling strengths $s$, with the WV fixed at $\theta=8\pi/9$,
and $\delta=\mu=0$.\label{fig:g2-ab-coralation}}
\end{figure}

To thoroughly analyze the properties of $g^{(2)}_{a,b}$, we first
compute the specific expectation values of $\langle\hat{a}^{\dagger}\hat{a}\hat{b}^{\dagger}\hat{b}\rangle$,
$\langle\hat{a}^{\dagger}\hat{a}\rangle$, and $\langle\hat{b}^{\dagger}\hat{b}\rangle$
for the state $\vert\Phi\rangle$. Given the lengthy derivations of these expressions, the detailed steps are provided in Appendix \ref{sec:A1}. In studies of two mode radiation fields, the quantity $\langle\hat{a}^{\dagger}\hat{a}\hat{b}^{\dagger}\hat{b}\rangle$ describes the intensity correlation between the two modes, while $\langle\hat{a}^{\dagger}\hat{a}\rangle$,
and $\langle\hat{b}^{\dagger}\hat{b}\rangle$ correspond to the mean photon numbers of the two modes \citep{Rivera-Dean_2026,PRXQuantum.5.040319}, respectively. Using this correlation function, we quantify the degree of photon correlation between different modes, where $g^{(2)}_{a,b}>1$ indicates positive photon correlation between modes a and b. Specifically, if $s=0$, $g^{(2)}_{a,b}$ reduces to the SOCC function of the initial ECSs, and is expressed as $g^{(2)}_{a,b,\phi}=0$. Figure \ref{fig:g2-ab-coralation} shows that for $s=0$, the pointer and measured system are uncoupled. The initial state (red curve in Fig. \ref{fig:g2-ab-coralation}) exhibits no correlation between modes in ECSs, indicating a vanishing coincidence probability. However, after applying a PVNMs, this conditional probability becomes nonzero. Consistent with previous analysis, it approaches $1$ for weak $r$ $(0<r<1)$, due to amplification by anomalous WVs. This behavior shows that post-selection redistributes the inter-mode intensity correlations of the ECS. To determine whether the same measurement process also modifies the genuinely quantum correlations between the two modes, we next analyze the Hillery-Zubairy and EPR entanglement criteria.

\section{The effects on entanglement\label{sec:5}}

In this section, we investigate the effects of PVNMs on the quantum correlations of ECSs. We examine the Hillery-Zubairy and EPR correlations of ECSs after measurement.

\subsection{Hillery-Zubairy correlation\label{subsec:5-1} }

\begin{figure}
\begin{centering}
\includegraphics[width=1\columnwidth]{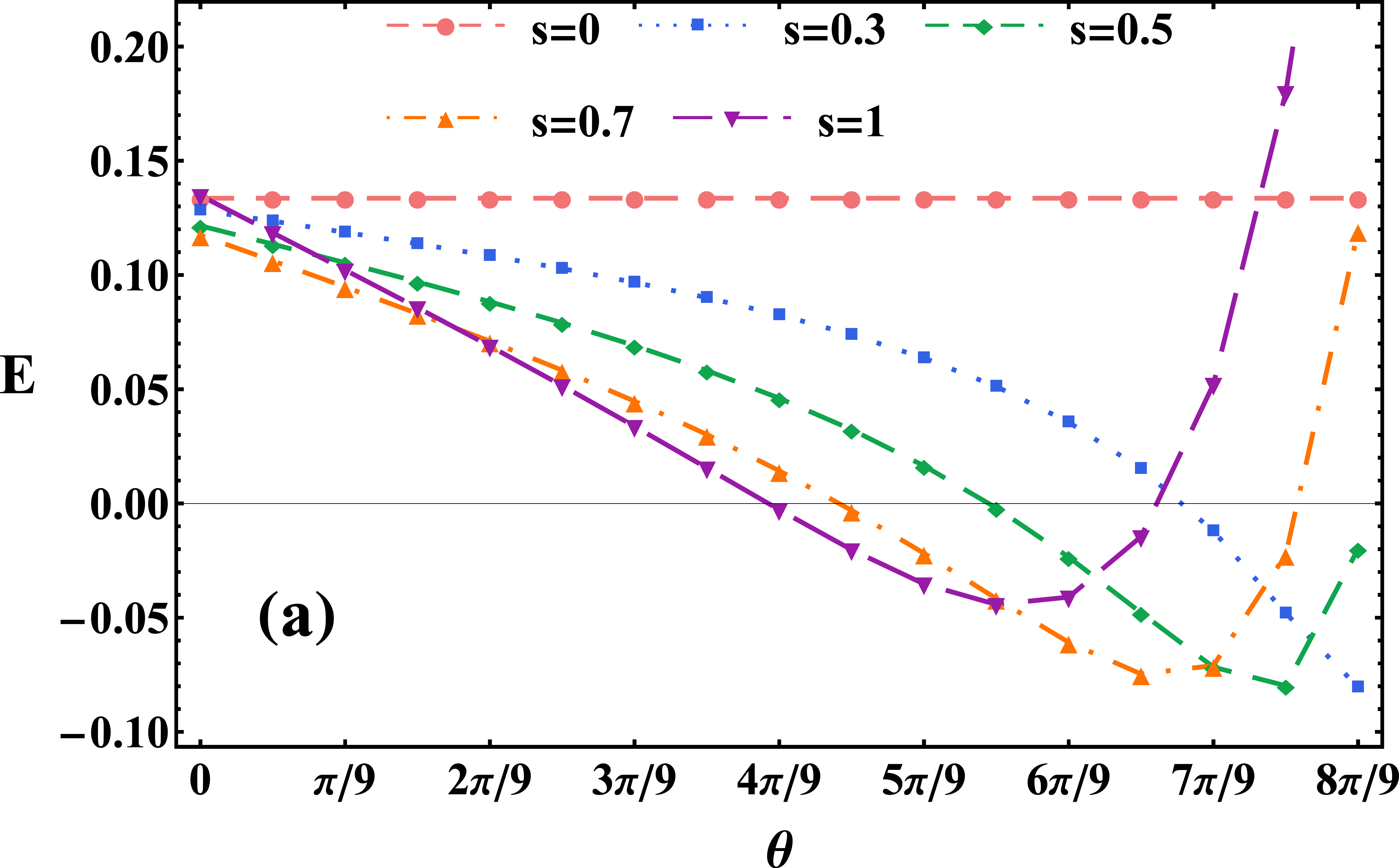}
\par\end{centering}
\begin{centering}
\includegraphics[width=1\columnwidth]{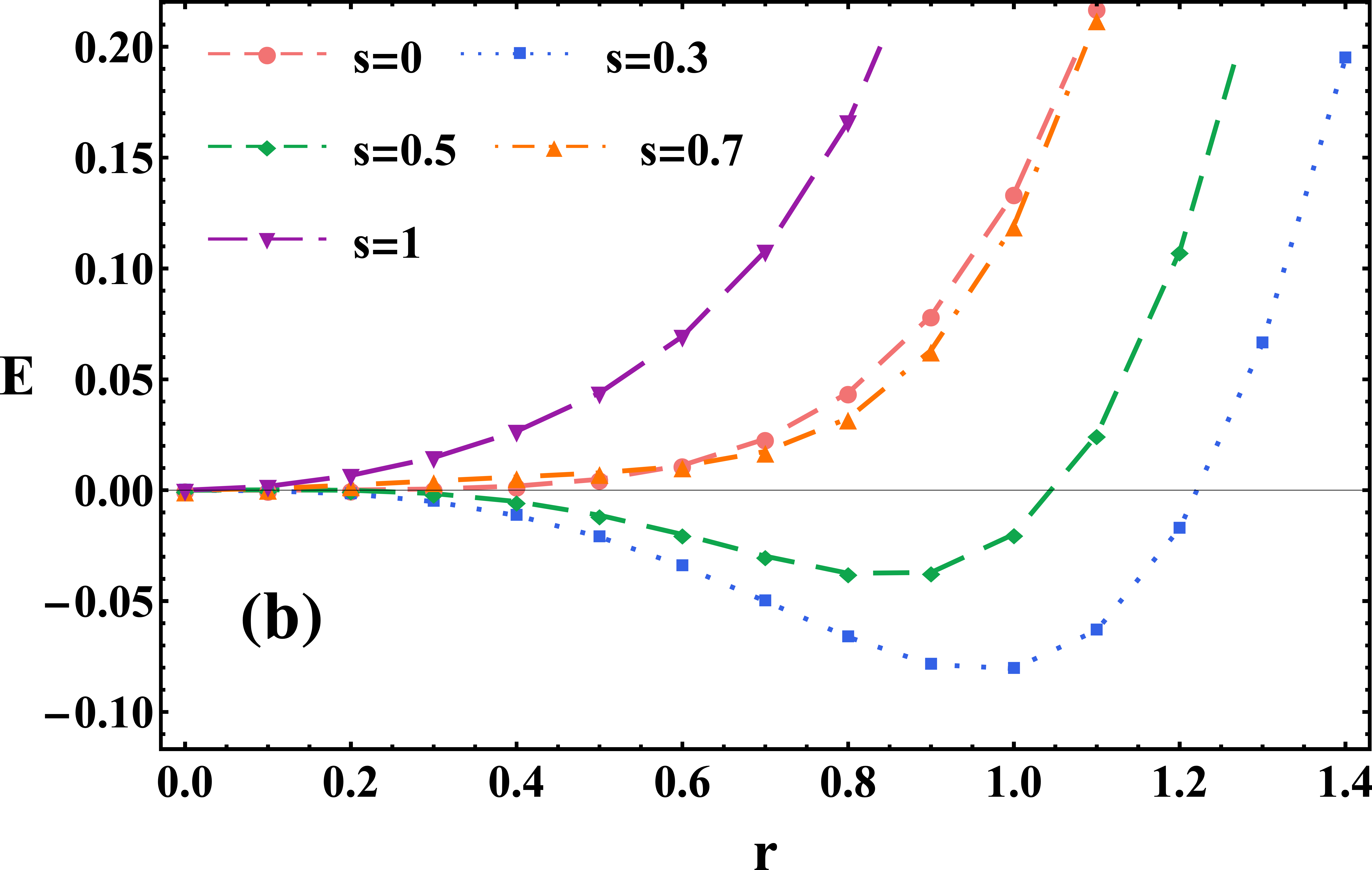}
\par\end{centering}
\caption{(a) $E$ as functions of the WVs parameter $\theta$ for different
coupling strengths $s$, with fixed $r=1$, (b) $E$ as functions
of $r$ for different coupling strengths $s$, with the WV
fixed at $\theta=8\pi/9$. Here, $\delta=\mu=0$.\label{fig:HZC}}
\end{figure}

The Hillery-Zubairy (HZ) correlation quantifies entanglement in two-mode quantum fields by comparing the difference \citep{PhysRevLett.96.050503,PhysRevLett.131.150201,297d-3tbj}.
The HZ correlation is in mathematical form, and its physical implications are characterized
as follows \citep{PhysRevA.74.032333}:

\begin{align}
E =\langle\hat{a}^{\dagger}\hat{a}\rangle\langle\hat{b}^{\dagger}\hat{b}\rangle-|\langle\hat{a}\hat{b}\rangle|^{2}.\label{eq:entanglement-Eq}
\end{align}

This entanglement criterion underscores the significance of the correlation
$\langle\hat{a}\hat{b}\rangle$ between the two modes of a state. Entanglement between the two modes arises when $E<0$. Additionally, the relation $\vert\langle\hat{a}\hat{b}\rangle\vert^{2}\le\langle\hat{a}^{\dagger}\hat{a}\rangle\langle\hat{b}\hat{b}^{\dagger}\rangle$
 holds \citep{PhysRevResearch.3.033095}, yielding the bound that the entanglement condition for two-mode fields is restricted to $-\langle a^{\dagger}a\rangle\le E<0.$ 
For the initial ECSs, $E_{\phi}=\mathcal{N}^{4}r^{4}>0$. 
We analyze compliance with the HZ correlation
for the final state $|\Phi \rangle$ by computing  Eq. (\ref{eq:entanglement-Eq}).
The explicit form of $E$ follows directly from the expectation values derived in Appendix \ref{sec:A1}. Figure \ref{fig:HZC} demonstrates WM control of HZ correlation entanglement. Figure \ref{fig:HZC}(a) maps $E$ against $\theta$ for varied $s$, identifying that the entanglement strength $E$ exhibits significant enhancement with the evolution of the WV parameter $\theta$ at a fixed $r=1$. Larger WVs demonstrate
an enhanced entanglement effect under the coupling coefficient $s=0.3$,
consistent with the earlier conclusion. Figure \ref{fig:HZC}(b) displays how $s$ modulates $E$ at a fixed WV $\langle\hat{\sigma}_{x}\rangle_{w}=5.671$,
with the weak coupling strength, which can exhibit an enhanced entanglement effect at $r=1$. The condition $\delta=\mu=0$ has been shown to enhance the entanglement between the two bosonic modes in the final state $|\Phi\rangle$ when compared to the ECSs. This enhancement
is significant when large WVs are present after a PVNMs. To illustrate this more clearly, we plot the evolution of $E$ versus the coupling strength parameter $s$ and the WV angle $\theta$, as shown in Figs. \ref{fig:HZC}(a) and (b), respectively. The entanglement behavior exhibits similarities across different values of $s$ and $\theta$, further indicating that appropriate coupling strength parameters $s$ and large anomalous WVs can induce more profound quantum effects compared to the initial state $|\phi\rangle$.

\subsection{Einstein--Podolsky--Rosen correlation\label{subsec:5-2}}

\begin{figure}
\begin{centering}
\includegraphics[width=1\columnwidth]{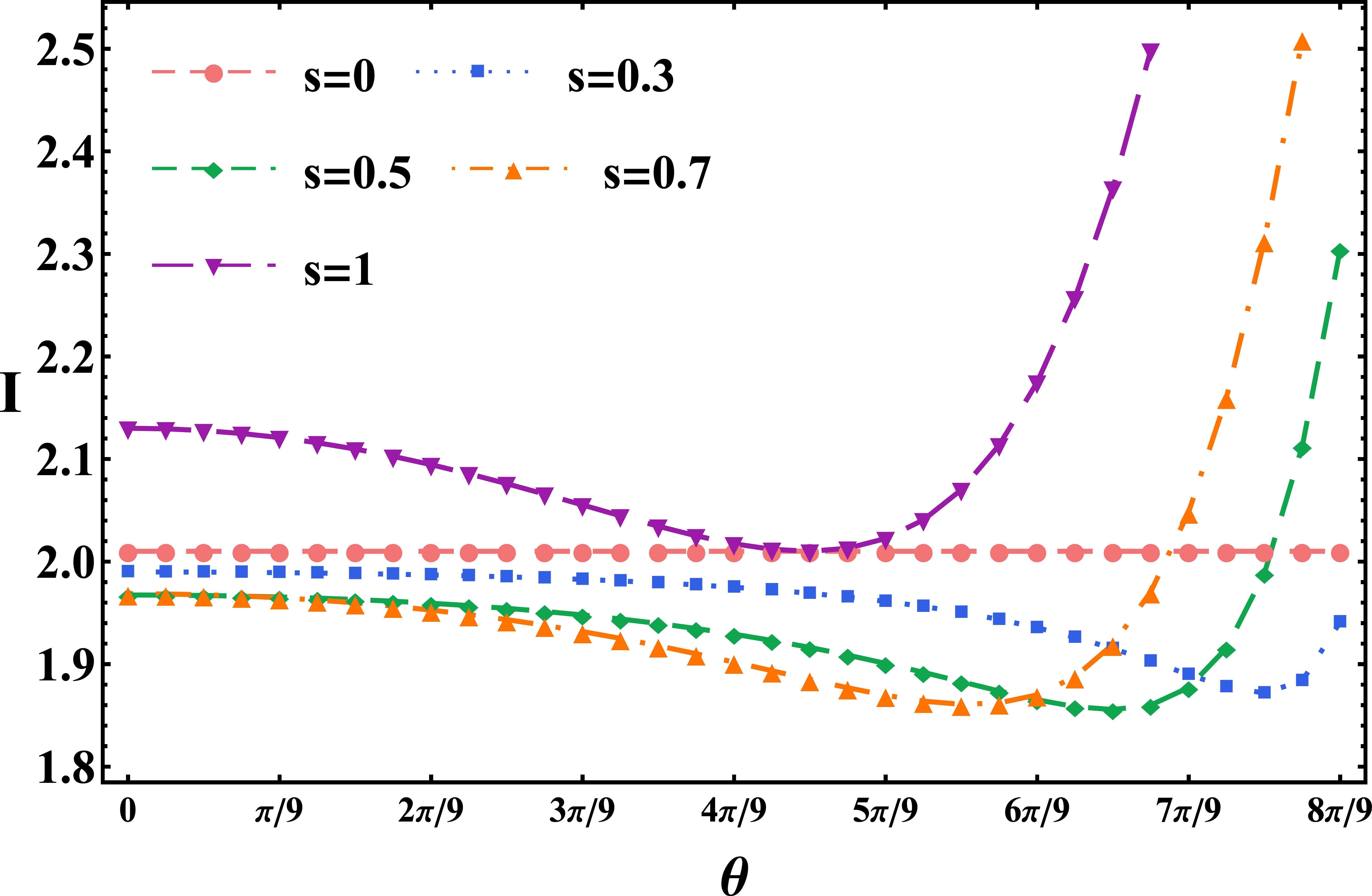}
\par\end{centering}
\caption{EPR correlations ($I$) as functions of $\theta$ for different coupling
strengths $s$, with the fixed at $r=0.1$ and $\delta=\mu=0$.\label{fig:EPR-correlations}}
\end{figure}

A novel inseparability criterion for continuous-variable systems, formulated in terms of the total variance of EPR-type operator pairs, has been introduced \citep{PhysRevA.80.012315}. For arbitrary continuous-variable states, violation of the bound provides a sufficient condition for inseparability. For the non-Gaussian ECSs considered here, failure to violate the bound does not imply separability.
Leveraging this criterion, we calculate the total variance of EPR-type operators for the ECSs under $|\Phi\rangle$ to assess the influence of
WM on the system's properties of inseparability. For two-mode systems, EPR correlations are explicitly expressed through the variances of the operators $\hat{X}_{1}-\hat{X}_{2}$ and $\hat{P}_{1}+\hat{P}_{2}$ \citep{REN2019106}, given by: 

\begin{align}
\text{I} & =\langle\Delta^{2}(\hat{X}_{1}-\hat{X}_{2})\rangle+\langle\Delta^{2}(\hat{P}_{1}+\hat{P}_{2})\rangle\nonumber \\
 & =2(1+\langle\hat{a}^{\dagger}\hat{a}\rangle+\langle\hat{b}^{\dagger}\hat{b}\rangle-\langle\hat{a}^{\dagger}\hat{b}^{\dagger}\rangle-\langle\hat{a}\hat{b}\rangle)\nonumber \\
 & \quad-2(\langle\hat{a}\rangle-\langle\hat{b}^{\dagger}\rangle)(\langle\hat{a}^{\dagger}\rangle-\langle\hat{b}\rangle).\label{eq:ERP}
\end{align}
where $\hat{X}_{1}=(\hat{a}+\hat{a}^{\dagger})/\sqrt{2},\;\hat{X}_{2}=(\hat{b}+\hat{b}^{\dagger})/\sqrt{2}$,
$\hat{P}_{1}=(\hat{a}-\hat{a}^{\dagger})/(i\sqrt{2})$ and $\text{\ensuremath{\hat{P}_{2}}}=(\hat{b}-\hat{b}^{\dagger})/(i\sqrt{2})$.
A violation of the separability bound, $I<2$, certifies two-mode
inseparability, whereas $I\geq2$ renders this second-moment criterion inconclusive \citep{PhysRevA.93.032330,PhysRevLett.84.2722}. To assess the influence of PVNMs on EPR correlations, we evaluate the requisite expectation values for $|\Phi\rangle$ (Appendix \ref{sec:A1}) and insert them into the EPR criterion, yielding a closed-form expression for the corresponding correlation measure.
If there is no coupling between the measuring device and the measured system ($s=0$), the EPR correlation function reduces to $I_{\phi}=2+r^{2}[\cos(2\mu)+\tanh(r^{2}/2)]>2$, corresponding to the ECSs. As shown in Fig. \ref{fig:EPR-correlations}, we systematically vary the coupling strength s while keeping $r$, $\mu$ and $\delta$ fixed, the flat red line ($s=0$) serves as a reference level $I>2$, By treating $I$ as a function of the WV angle $\theta$ and comparing different coupling strengths $s$ (with fixed $r=0.1$), the results similarly demonstrate that large WVs and moderate coupling enable the maximum reduction of $I$. For arbitrary parameters $r>0$ and non-minimal WV angles, the PVNMs can enhance the EPR correlation, thereby exceeding the entanglement of the original ECSs.

\section{Fidelity\label{sec:6}}

\begin{figure}[t]
\begin{centering}
\includegraphics[width=1\columnwidth]{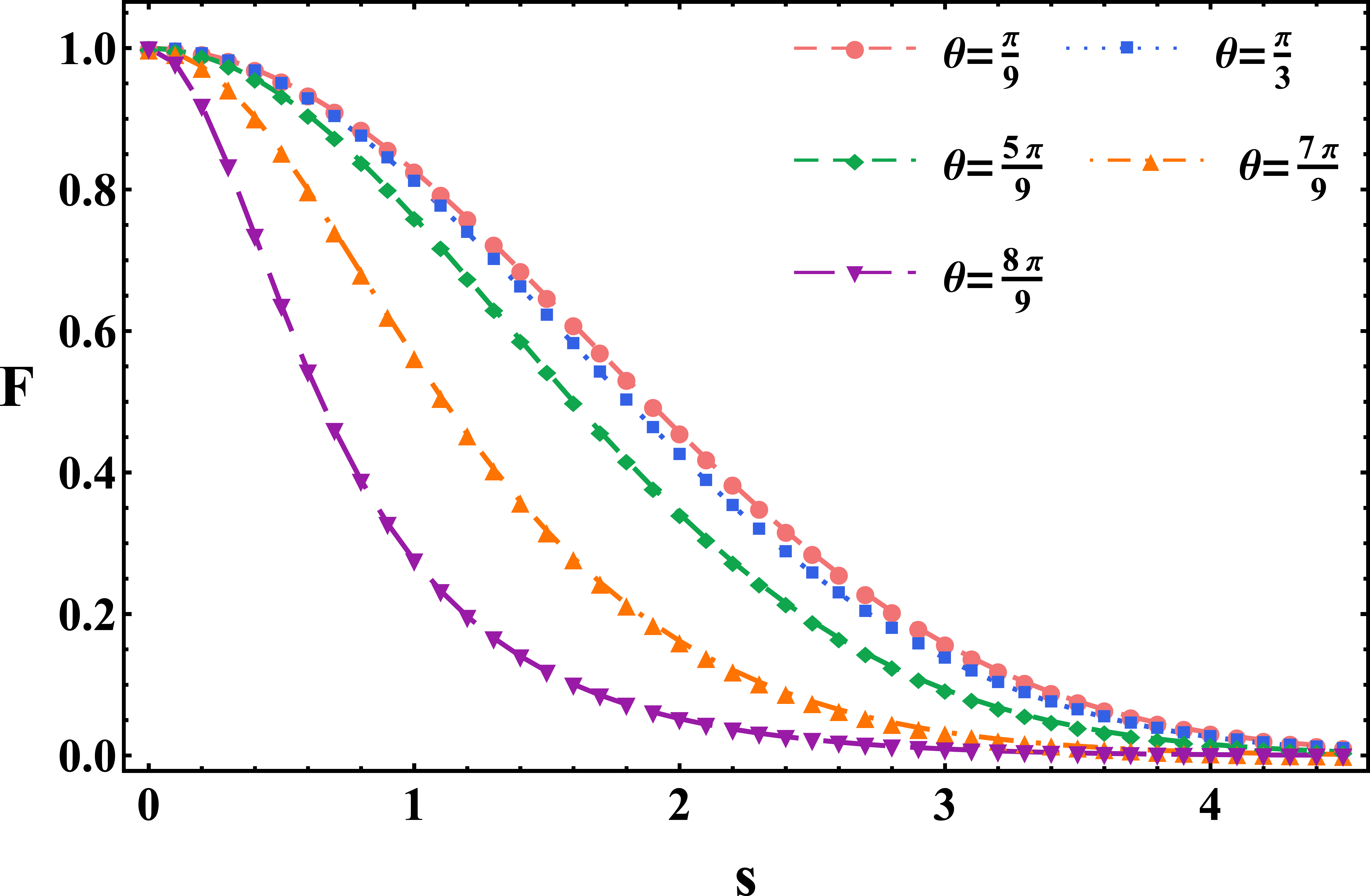}
\par\end{centering}
\caption{Fidelity $F$ as functions of $s$ for different WV parameter
$\theta$. Here, $r=0.1$ and $\delta=\mu=0$.\label{fig:fidelity}}
\end{figure}

To quantitatively assess the proximity between the initial pointer state  $\vert\phi\rangle$ and the final pointer state $\vert\Phi\rangle$, we adopt the fidelity for pure quantum states \citep{Jozsa01121994}. This function is defined as the squared modulus of their inner product, $F(\vert\phi\rangle,\vert\Phi\rangle)$=$\vert\langle\phi\vert\Phi\rangle\vert^{2}$, and is strictly bounded by $0\le F\le 1$. The limiting values correspond to orthogonal (perfectly distinguishable) and identical (fully overlapping) states, respectively. As an operationally meaningful quantifier of statistical overlap. Moreover, it directly relates to the distinguishability of the two preparations in measurement scenarios \citep{761271}. In the present analysis, we compute this fidelity as follows:

\begin{align}
F & =\vert\langle\phi\vert\Phi\rangle\vert^{2}=\frac{|\kappa|^{2}}{4}\vert[t_{+}I_{+}+t_{-}I_{-}]\vert^{2}.\label{fidelity}
\end{align}

Here, $\kappa$ represents the normalization coefficient, and with $I_{\pm}=\langle\phi\vert\hat{D}_a(\pm s/2)\vert\phi\rangle
=\mathcal{N}^2\{
(1+e^{-is\,\operatorname{Im}\alpha})e^{-s^2/8}
+e^{-|\alpha|^2/2}e^{-is\operatorname{Im}[\alpha]/2}
[
e^{-|\alpha-(s/2)|^2/2}
+e^{-|\alpha+(s/2)|^2/2}
]\}.
$
Figure \ref{fig:fidelity} presents an analysis of the $F$ as a function of $s$, demonstrating its exponential decay with respect to both the coupling strength $s$ and parameter $r$. This decay clearly illustrates the transformation of the initial even coherent state $\vert\phi\rangle$ via PVNMs. As depicted in Figure  \ref{fig:fidelity}, fidelity $F$ monotonically decreases with increasing $s$ across all WM conditions, progressively driving the initial and final states toward distinguishability. This trend remains robust regardless of the WV magnitude. However, for a fixed value of $s$, a large WV amplitude induces a more pronounced deviation in the evolved state $\vert\phi\rangle$, underscoring the enhanced back-action associated with large WVs. Crucially, the coupling parameter $s$ governs the crossover between the measurement regimes: in the limit $s\ll 1$, the $F$ asymptotically approaches unity (1), characteristic of a minimal disturbing weak measurement. However, for $s\gg 1$, $F$ tends to zero, corresponding to the strong projective limit where the states become fully distinguishable \citep{PhysRevA.103.052215,PhysRevA.102.062206}. This asymptotic behavior not only validates the theoretical framework but also offers profound insights into the controlled transition from weak to strong post-selected measurements and the resulting manipulation of quantum-state distinguishability.

For a sufficiently small coupling strength parameter $s\ll1$, the fidelity function in Eq. (\ref{fidelity}) simplifies to:

\begin{equation}
F \approx 4 \kappa^{\prime2}\{1+[s\langle\hat{\sigma}_{x}\rangle_{w}(1+e^{-\vert\alpha\vert^{2}})r\sin(\mu)]^{2}\},
\end{equation}
where 
$\kappa^{\prime}=1/\{2+8s\mathcal{N}^{2}(1+e^{2r})r\sin(\delta)\sin(\mu)\tan(\theta/2)+s^{2}\vert\langle\hat{\sigma}_{x}\rangle_{w}\vert^{2}[1-2\mathcal{N}^{2}(1+e^{2r}])r^{2}\cos(2\mu)]\}$,
a short interaction time and weak coupling mean that after the measurement,
the system and pointer stay nearly in the same state, with the difference
between $\vert\phi\rangle$ and $\vert\Phi\rangle$ approaching $1$. This indicates a WM, where the interaction is too weak to produce a significant dial offset from a single measurement. When performed over short timescales or with weak couplings, multiple consecutive measurements are required to achieve the desired statistical outcomes. Since large WVs in post-selective WM are associated with low post-selective probabilities, multiple measurements are essential to obtain statistically reliable results. The preceding analysis demonstrates that controlling the coupling strength parameter $s$ within a sufficiently small range is crucial for conducting post-selective WM. In this work, the enhancement of ECSs properties is achieved under these post-selective WM conditions.
In contrast, in the strongly coupled regime ($s\gg1$), the fidelity $F$ approaches zero, causing the system to lose all characteristics associated with the initial ECSs. This is because the state transitions to a completely different state. In post-selection, strong measurements render WVs irrelevant. In this case, the measured system is observed with respect to post-selective expectations \citep{Aharonov2008,YAharonov_1991}. In the developed scheme, the conditional expectation value of $\hat{\sigma}_{x}$ for the preselected state $|\psi_{i}\rangle$ and post-selected state $|\psi_{f}\rangle$ can be written as: 

\begin{equation}
\langle\hat{\sigma}_{x}\rangle_{c}=\sum_{j}a_{j}\frac{\vert\langle\psi_{f}\vert a_{j}\rangle\langle a_{j}\vert\psi_{i}\rangle\vert^{2}}{\sum_{i}\vert\langle\psi_{f}\vert a_{i}\rangle\langle a_{i}\vert\psi_{i}\rangle\vert^{2}}=\cos\delta\sin\theta.
\end{equation}

In this measurement mechanism, the initially prepared state of the measured system is eliminated after a single trial due to the strong coupling between the system and the pointer. Decreasing $\vert\phi\rangle$ and $\vert\Phi\rangle$ fidelity with coupling/WVs (Fig. \ref{fig:fidelity})
shows PVNMs alter the initial state, enabling state engineering.

\section{DISCUSSION\label{sec:7}}

The comprehensive characterization of the quantum features inherent to the output state, derived from the enhanced ECSs under post-selection, reveals that the PVNMs induce a thoughtful alteration of the initial state $|\psi\rangle$ into a qualitatively distinct final state $|\Phi\rangle$, with the deviation rigorously quantified by the fidelity. This post-selection framework,  recognized as a generalized measurement paradigm, provides a viable alternative for state optimization, as motivated in the introduction. 

\begin{figure}[t]
\begin{centering}
\includegraphics[width=1\columnwidth]{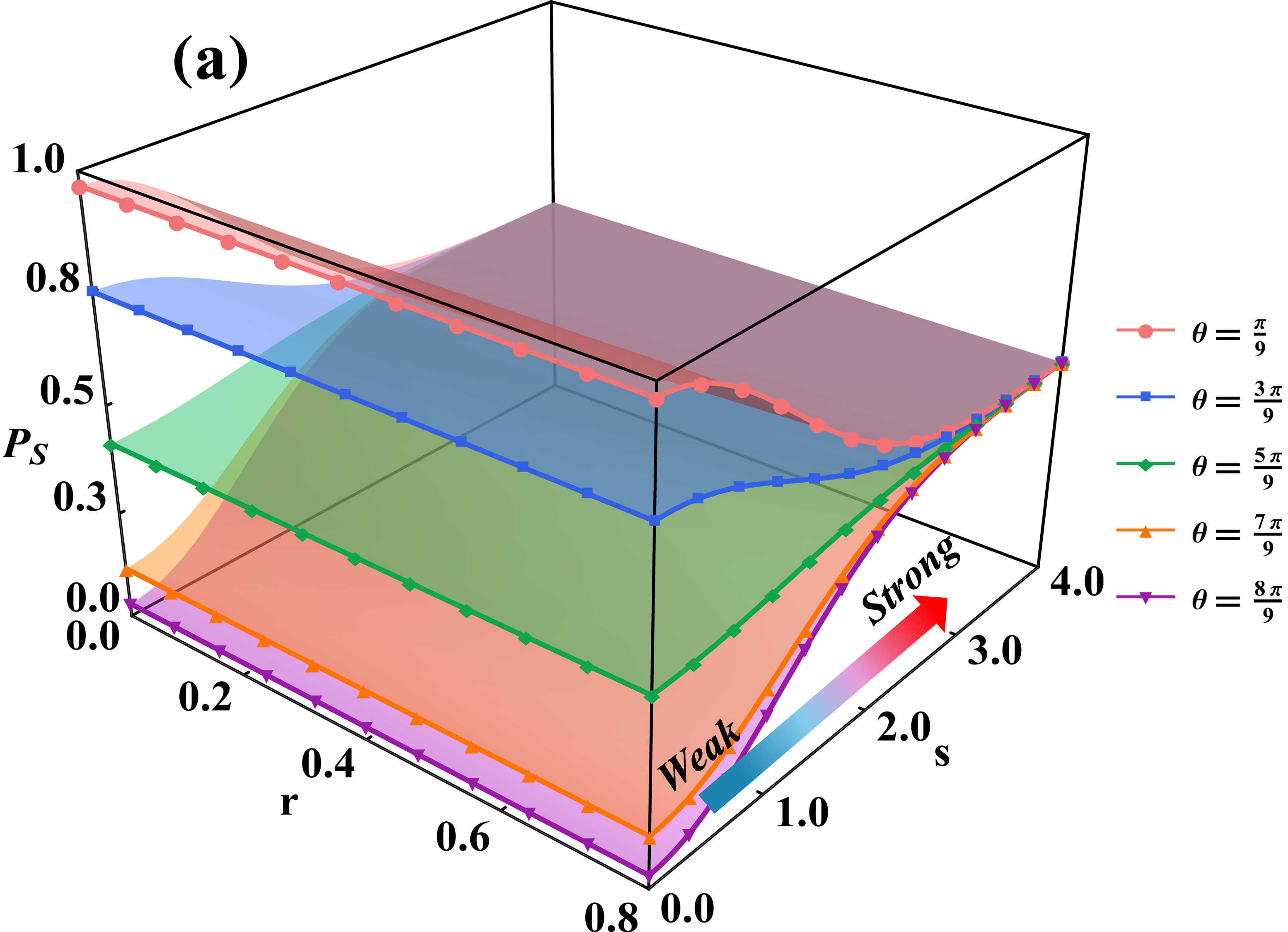}
\includegraphics[width=1\columnwidth]{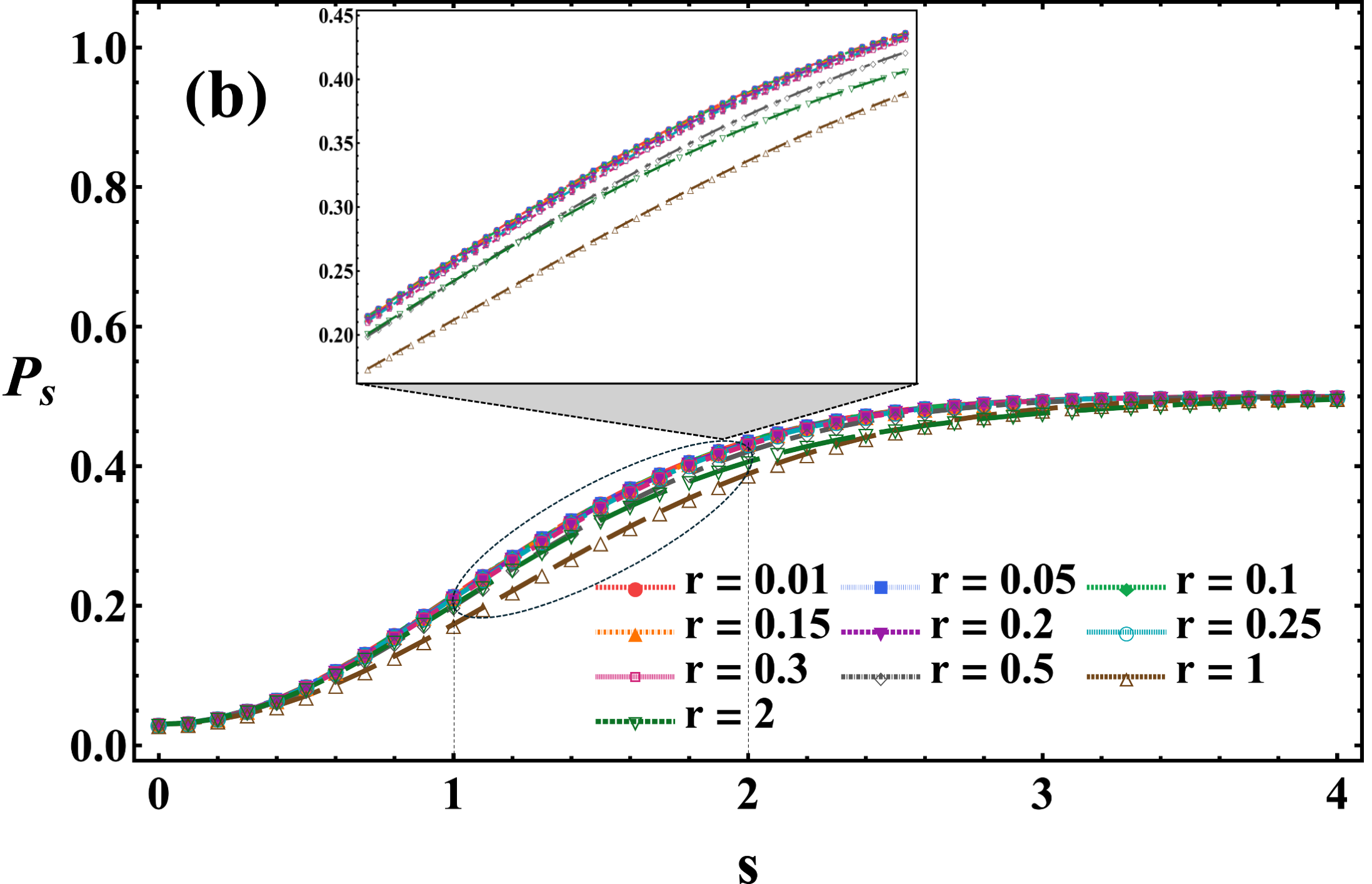}
\par\end{centering}
\caption{
The success probability $P_{\text{s}}$:
(a) as a function of $r$ and $s$ for different values of the WV
parameter $\theta$,
(b) as a function of $s$ for different values of $r$ at
$\theta=8\pi/9$.The inset curve displays the zoom-in curves of the assessed probability $P_{\text{s}}$ within the range $1 < s < 2$. The curve exhibits a clear linear increase across all $r$ values. Other parameters are $\mu=\delta=0$.\label{fig:success-probability}}
\end{figure}

Figure \ref{fig:success-probability} demonstrates that, in the WM regime, the success probability of the output state decreases with increasing WV and saturates as $s\geq 2.5$. Nevertheless, using an enhanced coupling strength $s$ facilitates the achievement of the final pointer state with a substantial success probability, even under significant anomalous WM conditions, which are typically characterized by reduced post-selection probabilities.

By systematically examining the dependence of fidelity on the coupling strength $s$ and the parameter $r$, we gain insights into measurement-induced back-action and the controllable weak-to-strong transition. Given these distinct parametric regimes, we propose six expected application scenarios (i.e., arbitrary values of $s$ and $r$,  strong coupling regime ($s\gg 1$) and arbitrary $r$,  strong coupling regime and small $r$ ($r\ll 1$),  weak coupling regime ($s\ll 1$) and arbitrary $r$, strong coupling regime and strong $r$, weak coupling regime and small $r$) each tailored to exploit the engineered distinguishability for advanced quantum state protocols. 

\noindent\textbf{Scenario I: Arbitrary values of $s$ and $r$}

In the general case, without imposing any approximation on the measurement strength $s$ or the coherent amplitude $r$, the post-selected pointer state takes the exact form:
\begin{align}
|\Phi\rangle
=&\frac{\kappa\mathcal N}{2} \Big\{
t_+\left[e^{-i\frac{s}{2}\operatorname{Im}\alpha}
|\alpha+\frac{s}{2}\rangle_a|0\rangle_b+|\frac{s}{2}\rangle_a|\alpha\rangle_b\right]
\nonumber \\
&+t_-\left[ e^{i\frac{s}{2}\operatorname{Im}\alpha} |\alpha-\frac{s}{2}\rangle_a|0\rangle_b + |-\frac{s}{2}\rangle_a|\alpha\rangle_b \right]
\Big\},
\end{align}
where, $|\alpha\pm (s/2),0\rangle=\hat{D}(\alpha\pm (s/2))\vert0,0\rangle$,
and the displacement operators $\hat{D}(\pm s/2)$ acting on mode-a.
The final state $|\Phi\rangle$ constitutes a superposition
of $|\alpha-(s/2),0\rangle$ and $|\alpha+(s/2),0\rangle$.
Considering the post-selection probability of the measured system,
which is $|\langle\psi_{f}|\psi_{i}\rangle|^{2}$, the post-selection success probability for this state is detailed in Eq. (\ref{eq:successful post-selection}).
This state is parameterized by both the WV and the coupling
strength $s$, featuring a detection probability $P_{\text{s}}$.


\noindent\textbf{Scenario II: Strong-coupling regime $s\gg 1$ and arbitrary $r$}

In this regime, no small $r$ approximation is required, and the complete coherent state structure of the ECSs is retained; the overlap satisfies  
$
\langle\phi|\hat D_a(s)|\phi\rangle\rightarrow0
$, for  $s \gg 1$. The normalized output state, therefore, becomes  
\begin{equation}
    \ket{\Phi}_{(1)} \approx \sqrt{2} [ c_{+}\hat{D}_{\text{a}}(s/2) + c_{-}\hat{D}_{\text{a}}^\dagger (s/2) ]\ket{\phi},
\end{equation}
where  
$
c_{\pm} =  \left[ \cos \left( \theta/2 \right) \pm e^{i\delta} \sin \left( \theta/2 \right) \right]/2.
$ 
Thus, the strong interaction generates two approximately orthogonal displaced ECSs branches, while the arbitrary parameter $r$ preserves the full coherent state amplitude. This state represents a measurement-induced multibranch entangled coherent superposition, and its post-selection success probability approaches $P_{\text{s}} \approx 1/2$.

\noindent\textbf{Scenario III: Strong coupling regime $s\gg 1$ and small parameter $r\ll 1$}

This scenario belongs to the regime of post-selected strong measurement.
As noted previously, WVs are no longer applicable here. The
system observable $\hat{\sigma}_{x}$ is given by the post-selected
expectation value $\langle\hat{\sigma}_{x}\rangle_{c}=\cos\delta\sin\theta$.
Correspondingly, Eq. (\ref{eq:Psi}) reduces to:
\begin{align}
\ket{\Phi}_{(2)} & \approx\frac{\kappa^{\prime}}{2\sqrt{2}}\Bigg[(t^{\prime}_{+}\vert\frac{s}{2}\rangle+t^{\prime}_{-}\vert-\frac{s}{2}\rangle)(2\vert0\rangle+\vert1\rangle)\nonumber \\
 & \quad+\alpha\bigg\{\big(t^{\prime}_{+}\vert\frac{s}{2},1\rangle+t^{\prime}_{-}\vert-\frac{s}{2},1\rangle\big)\vert0\rangle\bigg\},\label{eq:strong-weak-small-r}
\end{align}
The expression for $t^{\prime}_{\pm}=1\pm\langle\hat{\sigma}_{x}\rangle_{c}$,
and $\vert s/2,n\rangle$ for $n=0,1$, represents the displacement
Fock state of mode a. The normalization coefficient $\kappa^{\prime}$
is then given by
\begin{align}
\kappa^{\prime}  =\sqrt{2}/\left[1+\vert\langle\hat{\sigma_{x}}\rangle_{c}\vert^{2}+(1-\vert\langle\hat{\sigma}_{x}\rangle_{c}\vert^{2})\beta^{\prime}\right]^{1/2},
\end{align}
where $\beta^{\prime}=e^{-s^{2}/2}(e^{-2is\operatorname{Im}[\alpha]}+e^{-is\operatorname{Im}[\alpha]}+e^{is\operatorname{Im}[\alpha]}+1)/4$.
To explore entanglement and non-classical properties of the light
field in strong coupling ($s\gg1$) and small parameter ($r\ll1$)
regimes for quantum state preparation, we theoretically derived the
state vector using Eq. (\ref{eq:strong-weak-small-r}). Our derivation
demonstrated that the two modes of ECSs form an entangled state, creating
a superposition of displacements across distinct components (e.g.,
displaced Fock states $\vert s/2,0\rangle$ and $\vert-s/2,0\rangle$).
This state represents measurement-induced quantum state preparation,
exemplified by the Schrödinger cat state \citep{science.aaf2941}, in which displaced Fock states generate non-classical optical states via quantum superposition and exhibit multi-branch quantum features. This conversion of the light field into a superposition of two displaced states indicates potential applications in quantum information processing.

\noindent\textbf{Scenario IV: Weak coupling regime $s\ll 1$ and arbitrary $r$}

This situation falls under the category of PVNMs, where the measured value is determined by the WV. In this case, Eq. (\ref{eq:Psi}) simplifies to (non-normalized state):

\begin{equation}
\ket{\Phi}_{(3)} \approx \left[2+s \langle\hat{\sigma}_{x}\rangle_{w}(\hat{a}^{\dagger}-\hat{a})\right]\vert\phi\rangle.\label{eq:small-weak arbitrary-r}
\end{equation}

To elucidate the signal amplification mechanism in WM under weak coupling strength and arbitrary parameter $r$, we derived the pointer-state expression using post-selected WM theory.
Our derivation demonstrated that the state simplifies to the non-normalized form Eq. (\ref{eq:small-weak arbitrary-r}), in which the WV governs the measured value. This finding indicates that the second term predominates for anomalously large WVs, potentially facilitating quantum signal amplification and enhanced non-classical characteristics.

\noindent\textbf{Scenario V: Strong coupling regime $s\gg 1$ and Strong parameter $r\gg 1$}

In the simultaneous limits $s\gg1$ and $r\gg1$, both the intrinsic
overlap of the ECS components and the overlap between the measurement-induced displaced branches are exponentially suppressed,
such that $\mathcal{N}\rightarrow1/\sqrt{2}$ and
$\langle\phi|\hat{D}_{a}(s)|\phi\rangle\rightarrow0$. The conditional state, therefore, reduces to

\begin{align}
|\Phi\rangle_{(4)} \approx &
c_{+}\left[
e^{-is{\rm Im}[\alpha]/2}
\left|\alpha+\frac{s}{2},0\right\rangle
+
\left|\frac{s}{2},\alpha\right\rangle
\right]
\nonumber \\
&+
c_{-}\left[
e^{is{\rm Im}[\alpha]/2}
\left|\alpha-\frac{s}{2},0\right\rangle
+
\left|-\frac{s}{2},\alpha\right\rangle
\right],
\end{align}
where
$c_{\pm}=[\cos(\theta/2)\pm e^{i\delta}\sin(\theta/2)]/2$.
Hence, the strong interaction resolves the two measurement branches,
while the large coherent amplitude simultaneously renders the ECS
components nearly orthogonal. The resulting state is therefore a
coherent superposition of well-resolved two-mode branches, with
$P_{\rm suc}\rightarrow1/2$ and $F\rightarrow0$ in this asymptotic
limit.

\noindent\textbf{Scenario VI: Weak coupling regime $s\ll 1$ and small parameter $r\ll 1$}

This extreme scenario provides the post-measurement output state in (non-normalized state form):


\begin{align}
\ket{\Phi}_{(5)} \approx
& 2|0,0\rangle + \alpha(|1,0\rangle + |0,1\rangle) 
+ s\langle\hat{\sigma}_{x}\rangle_{w}
\Bigl[ |1,0\rangle
\nonumber \\
&+ \frac{\alpha}{2}
\bigl( \sqrt{2}\,|2,0\rangle - |0,0\rangle + |1,1\rangle \bigr)
\Bigr].
\end{align}

To investigate quantum state characteristics under weak coupling and small parameter regimes for quantum computing applications \citep{Y-PhysRevA.105.022608}, we theoretically derived the non-normalized quantum state $\vert\Phi\rangle_{(5)}$. Our derivation demonstrated that, owing to $r\ll1$, terms such as $\vert1\rangle\vert1\rangle$and $\vert0\rangle\vert1\rangle$are negligible. When the b mode resides in the vacuum state, the a mode displays a superposition of $\vert0\rangle$and $\vert1\rangle$ states; for sufficiently large WVs, a single-photon state in the a mode can be achieved. This indicates potential for quantum computing implementations \citep{4kpy-xfq2,rkzg-sdxn}.

\section{CONCLUSION AND OUTLOOK \label{sec:8}}

In summary, we have investigated the effects of PVNMs on the non-classical properties of ECSs. Using the exact conditional pointer state, we analyzed quadrature and sum squeezing, photon statistics, intermode correlations, HZ and EPR entanglement criteria, fidelity, and post-selection probability. The results show that anomalous WVs and moderate coupling can enhance the non-classical features of ECSs within finite parameter regions.

Post-selection enhances one quadrature while suppressing its conjugate counterpart, thereby strengthening sum squeezing. In the WM regime, the measured mode exhibits antibunching and sub-Poissonian statistics, while the unmeasured mode may display superbunching. Additionally, post-selection generates finite intermode intensity correlations and enhances entanglement signatures, as quantified by the HZ and EPR criteria. To clarify the measurement-induced state transformation, we analyze the fidelity between the initial and post-selected ECSs. 

The post-selection probability and its associated  limiting regimes characterize both the preparation probability and the parameter-dependent structure of the output states. These limiting cases naturally separate into six distinct scenarios, spanning arbitrary $s$ and $r$, strong coupling with arbitrary, small, and large $r$, and weak coupling with arbitrary and small $ r$ gap. Our study systematically builds on the WM framework to clarify the revolutionary role of post-selection in. Collectively, these scenarios describe the crossover from WV-dominated WM dynamics to well-resolved displaced-state superpositions, providing a compact guide to post-selected quantum-state engineering. Our results establish PVNMs as a tunable and probabilistic route for engineering ECSs. 

A realistic optical implementation could combine beam-splitter generation ECSs from coherent and squeezed-vacuum inputs with a
polarization-dependent weak interaction and post-selection, followed by balanced homodyne and photon-counting measurements of the resulting squeezing and photon statistics \citep{Fadrn2024}. Trapped-ion systems provide an alternative platform in which motional ECS preparation and single-ion weak measurements have both been experimentally demonstrated, increasing two-mode \citep{PhysRevA.109.032211}, while superconducting Kerr oscillators enable complementary preparation and control of entangled cat states \citep{Hoshi2025}. Future work will incorporate photon loss and dephasing,
 \citep{Ding2025,Wen2026}, finite detection efficiency, and imperfect post-selection \citep{Mei2026}. It will also identify experimentally
accessible parameter regimes in which the predicted non-classical
enhancement remains observable at finite post-selection probability
\citep{science.aaa2085,PhysRevA.110.062217,Yuanbek2026}.
\section{Methods}\label{sec:9}
In this work, all numerical calculations, analytical derivations, and data visualizations were executed using Wolfram Mathematica 15.0 and the Wolfram Quantum Framework, commercially licensed through the University of Chinese Academy of Sciences (UCAS). These tools provided high-precision capabilities critical for our quantum-mechanical analyses. All results underwent systematic cross-verification to ensure our numerical accuracy and logical reasoning.






\section*{Acknowledgments}
\textbf{J Yuanbek}, \textbf{ZA Bekele}, and \textbf{Z Liu} acknowledge the UCAS for providing access to use their commercially licensed Wolfram Mathematica 15.0 and the Wolfram Quantum Framework to professional computational environment.
This work was supported by the National Key Research and Development Program of China (Grant Nos. 2023YFB3611200, 2023YFB3609900), the National Natural Science Foundation of China (Grant Nos. 62474174, 62174157), the National Science and Technology Major Project (Grant No. 2025ZD0616200), Shenzhen Science and Technology Program (Grant No. JCYJ20241202123733043), the Science and Technology Project of Tianjin (Grant No. 24YFYSHZ00250).

\appendix

\section{\label{sec:A1}Explicit expectation values }

For completeness and rigor, this appendix provides the full derivations and closed-form expressions for all relevant quantities without approximation. These results are omitted from the main text because of their mathematical complexity. We first summarize the essential properties of coherent states in quantum optics.\citep{Descamps:26,t5c6-byhh}.

A coherent state $\vert\alpha\rangle$ is fundamentally characterized as an eigenstate of the annihilation operator $\hat{a}$, satisfying the eigenvalue equation $\hat{a}\vert\alpha\rangle=\alpha\vert\alpha\rangle$. Furthermore, the specific coherent states $\vert\pm\alpha\rangle$ are rigorously defined via the action of the displacement operator $\hat{D}(\alpha)=\exp [\alpha\hat{\alpha}^\dagger\ - \alpha^\ast\hat{a}]$ \citep{klimov2009group} on the vacuum state $\vert 0\rangle$, such that $\vert\alpha\rangle = \hat{D}(\alpha)\vert 0\rangle$. This foundational definition underpins the subsequent analytical expression and ensures the self-contained nature of the derivations provided in this appendix: 

\begin{widetext}

\begin{equation}
|\pm\alpha\rangle=e^{-\frac{|\alpha|^{2}}{2}}\sum^{\infty}_{n=0}(\pm1)^{n}\frac{\alpha^{n}}{\sqrt{n!}}|n\rangle,
\end{equation}

when the inner product is taken between a Fock state ($\vert n\rangle$)
and another coherent state$\vert\beta\rangle$, it yields $\langle n|\alpha\rangle=\langle0|\hat{a}^{n}|\alpha\rangle/\sqrt{n!}=\alpha^{n}\langle0|\alpha\rangle/\sqrt{n!}$,
and $\langle\beta|\alpha\rangle=\exp\left[(\beta^{*}\alpha-\beta\alpha^{*})/2\right]\exp[-|\beta-\alpha|^{2}/2]$ \citep{Fadrn2024,PhysRevA.110.033523}.

1. The expectation values of observables $\hat{a}$, $\hat{a}^{2}$and
$\hat{a}^{\dagger}\hat{a}$ in the eigenstates of ECSs ($\vert\Phi\rangle$)
are rigorously determined through closed-form analytical expressions.

\begin{align}
\langle a\rangle & =\frac{\vert\kappa\vert^{2}}{4}\bigg\{2\big(1+\vert\langle\hat{\sigma}_{x}\rangle_{w}\vert^{2}\big)\mathcal{N}^{2}\alpha(1+\exp[-\vert\alpha\vert^{2}])+2s \mathrm{Re}[\langle\hat{\sigma}_{x}\rangle_{w}]+\left(1+\langle\hat{\sigma}_{x}\rangle_{w}\right)^{\ast}\left(1-\langle\hat{\sigma}_{x}\rangle_{w}\right)\left(k_{11}-\frac{s}{2}\beta_{1}\right)\nonumber \\
 & \quad+\left(1-\langle\hat{\sigma}_{x}\rangle_{w}\right)^{\ast}\left(1+\langle\hat{\sigma}_{x}\rangle_{w}\right)\left(k_{12}+\frac{s}{2}\beta\right)\bigg\},
\\
\langle a^{2}\rangle & =\frac{\vert\kappa\vert^{2}}{4}\bigg\{2(1+\vert\langle\hat{\sigma}_{x}\rangle_{w}\vert^{2})(\mathcal{N}^{2}\alpha^{2}(1+\exp[-\vert\alpha\vert^{2}])+\frac{s^{2}}{4})+4s\mathrm{Re}[\langle\hat{\sigma}_{x}\rangle_{w}]\mathcal{N}^{2}\alpha(1+\exp[-\vert\alpha\vert^{2}])
\nonumber \\
 & \quad+\left(1+\langle\hat{\sigma}_{x}\rangle_{w}\right)^{\ast}\left(1-\langle\hat{\sigma}_{x}\rangle_{w}\right)(k_{21}-sk_{11}+\frac{s^{2}}{4}\beta_{1})+\left(1-\langle\hat{\sigma}_{x}\rangle_{w}\right)^{\ast}\left(1+\langle\hat{\sigma}_{x}\rangle_{w}\right)(k_{22}+sk_{12}+\frac{s^{2}}{4}\beta)\bigg\},
\\
\langle a^{\dagger}a\rangle & =\frac{\vert\kappa\vert^{2}}{4}\bigg\{2(1+\vert\langle\hat{\sigma}_{x}\rangle_{w}\vert^{2})(\mathcal{N}^{2}|\alpha|^{2}+\frac{s^{2}}{4})+4\mathrm{Re}[\langle\hat{\sigma}_{x}\rangle_{w}]s\mathcal{N}^{2}Re[\alpha](1+\exp[-\vert\alpha\vert^{2}])+(1+\langle\hat{\sigma}_{x}\rangle_{w})^{\ast}(1-\langle\hat{\sigma}_{x}\rangle_{w})\nonumber \\
 &\quad \times(k_{31}+(k_{11}-k_{41})\frac{s}{2}-\frac{s^{2}}{4}\beta_{1})+(1-\langle\hat{\sigma}_{x}\rangle_{w})^{\ast}(1+\langle\hat{\sigma}_{x}\rangle_{w})(k_{32}-(k_{12}-k_{42})\frac{s}{2}-\frac{s^{2}}{4}\beta)\bigg\},
\end{align}

where $\beta$ is defined in Sec. \ref{sec:2}, and  \begin{subequations}

\begin{align}
\beta_{1}  =\mathcal{N}^{2}\bigg\{\bigg(e^{2is\mathrm{Im}[\alpha]}+1\bigg)e^{-s^{2}/2}+e^{-\vert\alpha\vert^{2}/2}\bigg(e^{is\mathrm{Im}[\alpha]}e^{-|\alpha+s|^{2}/2}+e^{-is\mathrm{Im}[\alpha]}e^{-|s+\alpha|^{2}/2}\bigg)\bigg\},
\end{align}

\end{subequations}
analytical expressions for $K_{11}$ to $K_{42}$
are:

\begin{subequations}

\begin{align}
k_{11} & =\mathcal{N}^{2}\alpha(e^{2is\mathrm{Im}[\alpha]}e^{-s^{2}/2}+e^{-is\mathrm{Im}[\alpha]}e^{-|s+\alpha|^{2}/2}e^{-|\alpha|^{2}/2}),
\\
k_{12} & =\mathcal{N}^{2}\alpha(e^{-2is\mathrm{Im}[\alpha]}e^{-s^{2}/2}+e^{is\mathrm{Im}[\alpha]}e^{-|s-\alpha|^{2}/2}e^{-|\alpha|^{2}/2}),
\\
k_{21} & =\mathcal{N}^{2}\alpha^{2}(e^{2is\mathrm{Im}[\alpha]}e^{-s^{2}/2}+e^{-is\mathrm{Im}[\alpha]}e^{-|s+\alpha|^{2}/2}e^{-|\alpha|^{2}/2}),
\\
k_{22} & =\mathcal{N}^{2}\alpha^{2}(e^{-2is\mathrm{Im}[\alpha]}e^{-s^{2}/2}+e^{is\mathrm{Im}[\alpha]}e^{-|s-\alpha|^{2}/2}e^{-|\alpha|^{2}/2}),
\\
k_{31} & =\mathcal{N}^{2}|\alpha|^{2}e^{2is\mathrm{Im}[\alpha]}e^{-s^{2}/2},
\\
k_{32} & =\mathcal{N}^{2}|\alpha|^{2}e^{-2is\mathrm{Im}[\alpha]}e^{-s^{2}/2},
\\
k_{41} & =\mathcal{N}^{2}\alpha^{*}(e^{2is\mathrm{Im}[\alpha]}e^{-s^{2}/2}+e^{is\mathrm{Im}[\alpha]}e^{-|\alpha+s|^{2}/2}e^{-|\alpha|^{2}/2}),
\\
k_{42} & =\mathcal{N}^{2}\alpha^{*}(e^{-2is\mathrm{Im}[\alpha]}e^{-s^{2}/2}+e^{-is\mathrm{Im}[\alpha]}e^{-|\alpha-s|^{2}/2}e^{-|\alpha|^{2}/2}).
\end{align}

\end{subequations}

2. The expectation values of observables $\hat{b}$, $\hat{b}^{2}$and
$\hat{b}^{\dagger}\hat{b}$ in the eigenstates of ECSs ($\vert\Phi\rangle$)
are rigorously determined through closed-form analytical expressions.

\begin{align}
\langle\hat{b}\rangle & =\frac{\vert\kappa\vert^{2}}{4}\bigg\{2(1+\vert\langle\hat{\sigma}_{x}\rangle_{w}\vert^{2})\mathcal{N}^{2}\alpha(1+e^{-\vert\alpha\vert^{2}})+\left(1+\langle\hat{\sigma}_{x}\rangle_{w}\right)^{\ast}\left(1-\langle\hat{\sigma}_{x}\rangle_{w}\right)M_{11}+\left(1-\langle\hat{\sigma}_{x}\rangle_{w}\right)^{\ast}\left(1+\langle\hat{\sigma}_{x}\rangle_{w}\right)M_{12}\bigg\},
\\
\langle\hat{b}^{2}\rangle & =\frac{\vert\kappa\vert^{2}}{4}\bigg\{2(1+\vert\langle\hat{\sigma}_{x}\rangle_{w}\vert^{2})\mathcal{N}^{2}\alpha^{2}(1+e^{-\vert\alpha\vert^{2}})+\left(1+\langle\hat{\sigma}_{x}\rangle_{w}\right)^{\ast}\left(1-\langle\hat{\sigma}_{x}\rangle_{w}\right)M_{21}+\left(1-\langle\hat{\sigma}_{x}\rangle_{w}\right)^{\ast}\left(1+\langle\hat{\sigma}_{x}\rangle_{w}\right)M_{22}\bigg\},
\\
\langle\hat{b}^{\dagger}\hat{b}\rangle & =\frac{\vert\kappa\vert^{2}}{2}\mathcal{N}^{2}|\alpha|^{2}\bigg\{(1+\vert\langle\hat{\sigma}_{x}\rangle_{w}\vert^{2})+(1-\vert\langle\hat{\sigma}_{x}\rangle_{w}\vert^{2})e^{-s^{2}/2}\bigg\}.
\end{align}
Analytical expressions for $M_{11}$ to $M_{22}$ are:\begin{subequations}

\begin{align}
M_{11} & =\mathcal{N}^{2}\alpha(e^{is\mathrm{Im}[\alpha]}e^{-|\alpha+s|^{2}/2}e^{-|\alpha|^{2}/2}+e^{-s^{2}/2}),
\\
M_{12} & =\mathcal{N}^{2}\alpha(e^{-is\mathrm{Im}[\alpha]}e^{-|\alpha-s|^{2}/2}e^{-|\alpha|^{2}/2}+e^{-s^{2}/2}),
\\
M_{21} & =\mathcal{N}^{2}\alpha^{2}(e^{2is\mathrm{Im}[\alpha]}e^{-s^{2}/2}+e^{-is\mathrm{Im}[\alpha]}e^{-|s+\alpha|^{2}/2}e^{-|\alpha|^{2}/2}),
\\
M_{22} & =\mathcal{N}^{2}\alpha^{2}(e^{-2is\mathrm{Im}[\alpha]}e^{-s^{2}/2}+e^{is\mathrm{Im}[\alpha]}e^{-|s-\alpha|^{2}/2}e^{-|\alpha|^{2}/2}),
\end{align}

\end{subequations}

3. The expectation values of observables $\hat{a}\hat{b}$, $\hat{a}^{\dagger}\hat{b}$,
$\hat{a}^{\dagger}\hat{a}\hat{b}^{\dagger}\hat{b}$ and $\hat{a}^{2}\hat{b}^{2}$
in the eigenstates of ECSs are rigorously determined
through closed-form analytical expressions.

\begin{align}
\langle\hat{a}\hat{b}\rangle & =\frac{\vert\kappa\vert^{2}}{4}\bigg\{2s\mathrm{Re}[\langle\hat{\sigma}_{x}\rangle_{w}]\mathcal{N}^{2}\alpha(1+\exp[-\vert\alpha\vert^{2}])-(1+\langle\hat{\sigma}_{x}\rangle_{w})^{\ast}(1-\langle\hat{\sigma}_{x}\rangle_{w})\frac{s}{2}M_{11}
+(1-\langle\hat{\sigma}_{x}\rangle_{w})^{\ast}(1+\langle\hat{\sigma}_{x}\rangle_{w})\frac{s}{2}M_{12}\bigg\},
\\
\langle\hat{a}^{\dagger}\hat{b}\rangle & =\frac{\vert\kappa\vert^{2}}{4}\bigg\{2\big(1+\vert\langle\hat{\sigma}_{x}\rangle_{w}\vert^{2}\big)\mathcal{N}^{2}|\alpha|^{2}\exp[-\vert\alpha\vert^{2}]+2s\mathrm{Re}[\langle\hat{\sigma}_{x}\rangle_{w}]\mathcal{N}^{2}\alpha(1+\exp[-\vert\alpha\vert^{2}])
\nonumber \\
 & \quad+(1+\langle\hat{\sigma}_{x}\rangle_{w})^{\ast}(1-\langle\hat{\sigma}_{x}\rangle_{w})\left(K_{11}+\frac{s}{2}M_{11}\right)+(1-\langle\hat{\sigma}_{x}\rangle_{w})^{\ast}(1+\langle\hat{\sigma}_{x}\rangle_{w})\left(K_{12}-\frac{s}{2}M_{12}\right)\bigg\},
\\
\langle\hat{a}^{\dagger}\hat{a}\hat{b}^{\dagger}\hat{b}\rangle & =\frac{\vert\kappa\vert^{2}s^{2}}{8}\bigg\{(1+\vert\langle\hat{\sigma}_{x}\rangle_{w}\vert^{2})\mathcal{N}^{2}|\alpha|^{2}+\frac{1}{2}\big[\left(1+\langle\hat{\sigma}_{x}\rangle_{w}\right)^{\ast}\left(1-\langle\hat{\sigma}_{x}\rangle_{w}\right)M_{3}-\left(1-\langle\hat{\sigma}_{x}\rangle_{w}\right)^{\ast}\left(1+\langle\hat{\sigma}_{x}\rangle_{w}\right)M^{\dagger}_{3}\big]\bigg\},
\\
\langle\hat{a}^{2}\hat{b}^{2}\rangle & =\frac{\vert\kappa\vert^{2}s^{2}}{8}\bigg\{\big(1+|\langle\hat{\sigma}_{x}\rangle_{w}\vert^{2}\big)\mathcal{N}^{2}\alpha^{2}(1+\exp[-\vert\alpha\vert^{2}])
\nonumber \\
 & \quad +\frac{1}{2}\big[\left(1+\langle\hat{\sigma}_{x}\rangle_{w}\right)^{\ast}\left(1-\langle\hat{\sigma}_{x}\rangle_{w}\right)M_{21}+\frac{s^{2}}{4}\left(1-\langle\hat{\sigma}_{x}\rangle_{w}\right)^{\ast}\left(1+\langle\hat{\sigma}_{x}\rangle_{w}\right)M_{22}\big]\bigg\},
\end{align}
where
\begin{subequations}
\begin{align}
M_{3}&=\mathcal{N}^{2}\alpha^{2}(1+e^{-|\alpha|^{2}/2})e^{-2is\mathrm{Im}[\alpha]}e^{-s^{2}/2},
\\
K_{11} & =\mathcal{N}^{2}|\alpha|^{2}e^{is\mathrm{Im}[\alpha]}e^{-|\alpha+s|^{2}/2}e^{-|\alpha|^{2}/2},
\\
K_{12} & =\mathcal{N}^{2}|\alpha|^{2}e^{-is\mathrm{Im}[\alpha]}e^{-|\alpha-s|^{2}/2}e^{-|\alpha|^{2}/2}.
\end{align}
Appendix ends here.
\end{subequations}

\removewidetextbottomrule
\end{widetext}

\bibliography{ref-ECS}

\end{document}